\documentclass[12pt,preprint]{aastex}

\usepackage{epsfig}
\usepackage{times}
\usepackage{color,colordvi}

\def \deeg {$^\circ$}
\def \kms {${\rm km~s}^{-1}$}
\def \cc {${\rm ~cm}^{-3}$}
\def\HI{H$^\mathrm{o}$}

\def\lya{Ly$\alpha$}

\def\CaII{Ca$^\mathrm{+}$}
\def\HII{H$^\mathrm{+}$}

\def\HeI{He$^\mathrm{o}$}
\def\HeII{He$^\mathrm{+}$}
\def\FeII{Fe$^\mathrm{+}$}
\def\MgII{Mg$^\mathrm{+}$}

\def\chisq{$\chi ^ 2$}
\def\microG{$\mu$G}
\newcommand{\velinf} {V$_{\mathrm{ISM},\infty}$}
\newcommand{\veloneau}{V$_\mathrm{ISM,1AU}$}
\newcommand{\laminf} {$\lambda_{\mathrm{ISM}, \infty} $}
\newcommand{\lampeak}{$\lambda_\mathrm{peak}$}
\newcommand{\lamobs}{$\lambda_{Obs}$}
\newcommand{\betinf}{$\beta_{\mathrm{ISM}, \infty}$}
\newcommand{\teminf}{T$_{\mathrm{ISM}, \infty}$}
\newcommand{\theinf}{$\theta_{\infty}$}

\slugcomment{Astrophysical Journal, in press, \today}
\shorttitle{Correcting the record regarding the analyses of IBEX and STEREO data}
\shortauthors{Frisch et al.}

\begin{document}

\title{CORRECTING THE RECORD ON THE ANALYSIS OF IBEX AND STEREO DATA
REGARDING VARIATIONS IN THE NEUTRAL INTERSTELLAR WIND}

\author{P. C. Frisch}
\affil{Dept. Astronomy and Astrophysics, University of Chicago, Chicago, IL  60637}
\author{M. Bzowski} 
\affil{Space Research Centre of the Polish Academy of Sciences, Warsaw, Poland}
\author{C. Drews} 
\affil{Institute for Experimental and Applied Physics, Christian-Albrechts-University zu Kiel, Germany}
\author{T. Leonard} 
\affil{Space Science Center, University of New Hampshire, Durham, NH 03824, USA}
\author{G. {Livadiotis}} 
\affil{Southwest Research Institute, San Antonio, TX 78249, USA}
\author{D.~J. McComas} 
\affil{Southwest Research Institute, San Antonio, TX 78249, USA}
\altaffiltext{1}{University of Texas in San Antonio, San Antonio, TX 78249, USA}
\author{E. M{\"o}bius}
\affil{Space Science Center, University of New Hampshire, Durham, NH 03824, USA}
\author{N. Schwadron}
\affil{Space Science Center, University of New Hampshire, Durham, NH 03824, USA}
\author{ J. M. Sok{\'o}{\l}} 
\affil{Space Research Centre of the Polish Academy of Sciences, Warsaw, Poland}
\pagebreak
\begin{abstract}
The journey of the Sun through space carries the solar system through
a dynamic interstellar environment that is presently characterized by
a Mach $\sim 1$ motion between the heliosphere and the surrounding
warm partially ionized interstellar cloud.  The interaction between
the heliosphere and interstellar medium is an evolving process due to
variable solar wind properties and the turbulent nature of the
interstellar cloud that surrounds the heliosphere.  Frisch et
al. presented a meta-analysis of the
historical data on the interstellar wind flowing through the
heliosphere and concluded that temporal changes in the ecliptic
longitude of the flow direction with time were statistically indicated
by the data available in the refereed literature at the time of that
writing.  Lallement \& Bertaux disagree with
this result, and suggested, for instance, that a key instrumental
response function of IBEX-Lo was incorrect and that the STEREO
pickup ion data are unsuitable for diagnosing the flow of interstellar
neutrals through the heliosphere.  In this paper we first show that
temporal variations in the interstellar wind through the heliosphere
are consistent with our knowledge of the very local interstellar
medium.  The statistical analysis of the helium wind data is
revisited, and a recent correction of a typographical error in the
literature is incorporated into the new fits.  With this correction,
and including no newer IBEX results, these combined data still
indicate that a change in the longitude of the interstellar neutral
wind of $\lambda = 5.6^\circ \pm2.4^\circ $ over the past forty years
remains statistically likely, but an constant flow longitude is now
statistically possible.  Other scenarios for the selection of subsets
of these data used in the fitting process produce similar conclusions.
We show that the speculations made by Lallement \& Bertin about the
IBEX instrumental response function are incorrect, and that their
other objections to the data used in the meta-analysis are either
incorrect or unproven.  Further investigations of the historical
interstellar wind data and continuing analysis of additional IBEX data
may provide a more definitive result on the stability of the flow of
interstellar gas through the heliosphere.
\end{abstract}

\keywords{ISM: magnetic fields, heliosphere--- solar system:  general}


\section{Introduction} \label{sec:intro}

\nocite{Frisch:2013sci}

The solar system is traveling through a dynamically evolving
interstellar medium that leads to variations in the interstellar wind
through the solar system over geologically short time-scales
\citep{Frisch:2011araa,FrischMueller:2011ssr}.  The Local Interstellar
Cloud (LIC) now around the heliosphere is a low density and partially
ionized cloud \citep{SlavinFrisch:2008}.  The relative motions of the
Sun and LIC creates a wind of interstellar gas and dust through the
heliosphere whose velocity was first defined by measurements of the
fluorescence of solar 584\AA\ photons off of interstellar \HeI\ in the
gravitational focusing cone downwind of the Sun
\citep{WellerMeier:1974}.  The absence of interstellar \HI\ at the LIC
velocity toward the two nearby stars $\alpha$ Cen (1.3 pc) and 36 Oph
(6.0 pc) indicate that the Sun will emerge from the LIC in the
upwind\footnote{The terms ``upwind'' and ``downwind'' are referred to
  the direction of the heliosphere nose, which in turn is defined by
  the direction of the interstellar \HeI\ wind that flows into the
  inner heliosphere.}  direction anytime within the next 3,500 years
\citep{Lallementetal:1995,Wood36Oph:2000}, and that it entered the LIC
during the past 10,000 years \citep{Frisch:1994sci}.  Variations in
the interstellar wind direction over historical timescales, of about
40 years for the space age, suggest that interstellar neutrals, and
the pickup ions that form from them after ionization, provide the best
diagnostics of the longitude of the interstellar wind flowing through
the heliosphere.  The dynamic characteristics of the surrounding LIC
suggest that short-term variations in the interstellar wind direction
may exist.

In the paper titled "Decades-Long Changes of the Interstellar Wind
Through Our Solar System", Frisch et al.~(2013, hereafter Paper I)
have presented a meta-analysis of historical observations of the
interstellar \HeI\ wind data and found that the data indicated a
statistically likely shift in the wind longitude of $\sim 6.8^\circ$
over the past forty years, and that a constant flow longitude is
statistically unlikely.  The historical data set included in situ
measurements of the helium wind by IBEX
\citep{McComas:2012bow,Moebius:2012isn,Bzowski:2012isn} and Ulysses
\citep{Witte:2004}, measurements of pickup ions in the downwind
gravitational focusing cone by the ACE-SWICS, STEREO-PLASTIC, and
MESSENGER-FIPS \citep{Gloeckler:2004,Drews:2012,Gershman:2013}, the
upwind crescent by STEREO-PLASTIC \citep{Drews:2012}, and measurements
of the backscattered radiation from solar \HeI\ 584\AA\ emissions
fluorescing from interstellar \HeI\ in the inner heliosphere
\citep{WellerMeier:1974,Ajello:1978,Ajello:1979,WellerMeier:1981,DalaudierBertaux:1984,Vallerga:2004,Nakagawa:2008nozomi}.
This was the first meta-analysis that searched for possible historical
variations in the direction of the wind of heavy interstellar neutral
atoms through the heliosphere.

The original study in Paper I was restricted to heavy atoms and did
not include longitude information from the velocity vectors of either
interstellar \HI\ or interstellar dust grains flowing through the
heliosphere.  Both sets of particles are strongly filtered on a
time-variable basis, because of the solar magnetic activity cycle,
throughout the heliosphere and heliosheath regions
\citep{RipkenFahr:1983,RucinskiBzowski:1996,Saul:2013isn,Groganetal:1996zodi,Landgraf:2000,SlavinFrisch:2012,SterkenGruen:2012,StrubKrueger:2014}.

In a paper "On the Decades-Long Stability of the Interstellar Wind
through the Solar System", \citet[][hereafter LB14]{LB14} disagree
with the conclusions in Paper I.  The different viewpoints presented
in Paper I and LB14 fall into two categories, the most interesting of
which is the underlying science question that is encapsulated by the
differences of the two titles "On the "Decades-Long Changes of the
Interstellar Wind Through Our Solar System" compared to "Decades-Long
Stability of the Interstellar Wind through the Solar System".  The
second area of disagreement centers on the selection of data used in
the analysis, and the IBEX instrumental parameters.  LB14 favor the
interstellar \HeI\ flow direction determined in \citet{Moebius:2004},
where measurements of the interstellar \HeI\ wind found by Ulysses,
ACE, EUVE, and Prognoz 6 were combined to find the best-fitting
direction for the interstellar wind.  The groundwork for understanding
the interstellar helium wind is presented in the set of papers that
accompany \citet{Moebius:2004}.\footnote{The wind direction in this
  study needs to be corrected by 0.7\deeg\ to account for an
  unrecognized difference in coordinate epochs in the contributing
  data that led to an error of about 0.7\deeg\ in the wind direction
  \citep{Frischetal:2009ibex,Lallement:2010soho}.}

In this paper we first revisit the conclusions of Paper I, with the
proviso that new data on, and better modeling of, the interstellar
wind are creating a rapidly changing scientific data base upon which
our conclusions rely.  Data showing that the interstellar wind
\emph{could} change over historical times are discussed, and these
data are consistent with inhomogeneities in the LIC over spatial
scales comparable to the heliosphere dimensions (\S \ref{sec:scale}).
New statistical analyses are performed on the historical data (\S
\ref{sec:refit}), utilizing a new published correction for the
uncertainties on the NOZOMI interstellar \HeI\ wind direction (\S
\ref{sec:584n}).  A temporal change in the wind longitude is found to
be statistically likely, but a constant wind direction also becomes
statistically acceptable although less likely than a variable
direction.  LB14 objected to the use of the IBEX-Lo data in recovering
a longitude for the \HeI\ flow that differed from the value found by
Ulysses.  Methods for retrieving the flow longitude from IBEX-Lo data
are discussed in \S \ref{sec:ibextrajectory}, with further details on
the instrumental response function and properties in \S
\ref{sec:ibexirf} and Appendix \ref{app:ibex}.  Cloud temperature
plays an important role in the uncertainties of the parameter space
constrained by the IBEX-Lo data; the interstellar picture of the LIC
temperature is discussed in Appendix \ref{app:lic}.  The LB14 comments
on the STEREO data are discussed in \S \ref{sec:stereo}, where new
data are mentioned that contradict the hypothesis of significant
azimuthal transport of pickup ions inside of 1 AU, and in Appendix
\ref{app:stereo}.

\section{Possible spatial inhomogeneities in LIC}\label{sec:scale}

\subsection{Turbulence and edge effects }

The Sun moves through the surrounding interstellar cloud at a relative
velocity of $\sim 5$ AU per year, so that the 40-year historical
record of the interstellar wind velocity sampled interstellar scale
lengths of $\sim 200$ AU.  There is no ad hoc reason that the LIC
should be homogeneous and isotropic over such small spatial scales.
The mean free path of a thermal population of LIC atoms is $\sim 330$
AU,\footnote{The mean free path in the LIC is dominated by
charge-exchange and induced dipole scattering between \HI\ and
\HII\ atoms \citep{SpanglerSavage:2011cex}.}  which is larger than
the distance traveled through the LIC during the past 40 years.
Collisional coupling between atoms will break down over scales smaller
than the mean free path, allowing the formation of eddies that perturb
the gas velocity over heliosphere scale-lengths.  Paper I suggested
that a shift in the \HeI\ flow direction could be due to the
nonthermal turbulence of eddy motions.

A second source of small-scale turbulence in the LIC is Alfvenic
turbulence.  The LIC interstellar magnetic field (ISMF)
\citep{McComas:2009sci,Funsten:2013} affects clumping of the charged
particles over scales smaller than the collisional mean free path.
The LIC magnetic field strength is $\sim 3$ \microG, based on the
plasma pressure of energetic neutral atoms (ENAs) in the upwind direction and the magnetic
distortion of the heliotail \citep{Schwadronetal:2011sep}, and the
equipartition of energy between magnetic and thermal energies in the
LIC \citep{SlavinFrisch:2008}.  For field strengths of 3 \microG\ and
larger, the Lorentz force will couple protons tightly to the ISMF,
over $\sim 375$ km scales for a gas temperature of 7000 K.  This tight
coupling acts as a conduit for magnetic turbulence into the LIC that
affects the neutral population observed by IBEX through charge
exchange, and would perturb the neutral particle velocities over
spatial scales smaller than the collisional scale.

The location of the LIC in a decelerating group of clouds that appears
to be a fragment of the Loop I superbubble shell
\citep{Frisch:2011araa} would promote turbulence in the local ISMF.
If the surrounding decelerating flow of ISM
\citep{FGW:2002,Frisch:2011araa} consists of clouds in contact with
each other, then macro-turbulence may form in the LIC at the acoustic
velocity of about 8 \kms.  The angle between the LIC ISMF and the flow
of LIC gas through the local standard rest (LSR) is $86^\circ \pm
14^\circ$, indicating that the velocity and magnetic field vectors are
perpendicular \citep{FrischSchwadron:2014}.  Depending on the relative
thermal and magnetic pressures, compressive turbulence may form.
Magnetic turbulence may propagate along the bridge of polarized dust
that extends from near the Sun out to the North Polar Spur region
\citep{Frisch:2014icns}.  Magnetic disturbances will propagate along
the ISMF at the Alfven speed of $\sim 25$ \kms\ \citep[using an
  electron density of 0.07 \cc,][]{SlavinFrisch:2008}, organizing
additional disruption of particle velocities over scale sizes smaller
than the mean free path.

It is important to note that the thermal distribution of the
\HeI\ atoms smooths the time-interval sampled by the population
observed at 1 AU because of the different propagation times of fast
and slow atoms through the heliosphere.  A Maxwellian gas at 7000 K
has a Doppler width of $\sim 5$ \kms, which is a significant
percentage of the bulk LIC motion, causing faster particles to catch
up to slower particles in the incoming \HeI\ flow.  The Maxwellian
distribution will not apply over microscales less than the
330 AU mean free path of charge-exchange, and in principle the mean
free paths of the higher energy atoms will be larger than 330 AU
anyway.

Upper limits on an interstellar \lya\ component at the LIC velocity
toward 36 Oph, in the upwind direction, show that the Sun is near the
LIC edge \citep{Wood36Oph:2000}.  The high level of LIC ionization
that is inferred from photoionization models requires that $\sim
22\% ,~ 39\% , ~ 80\%$, and 19\% of the hydrogen, helium,
neon, and oxygen atoms at the heliosphere boundaries are ionized
\citep[][SF08]{SlavinFrisch:2008}.  A thin conductive interface
between the warm LIC and hot plasma of the Local Bubble maintains
these ionization levels (SF08).  A sharp density decrease and velocity
increase is expected at the edge between the LIC and conductive
interface.  The unknown distance to the upwind edge of the LIC allows
that such an interface is close to the heliosphere. However, no direct
observational evidence of this interface is yet found.

\subsection{Supersonic collisions between clouds} \label{sec:shock}

Disturbances due to the propagation of shocks through the cloud are
also possible sources of kinematical structure in the LIC.  The LIC
belongs to a decelerating flow of ISM through space that may create
dynamical interfaces in the interaction regions between clouds in
physical contact \citep{Frisch:2011araa}.  If the LIC is in physical
contact with either the Blue Cloud (BC) in the downwind direction, or
the G-Cloud (GC) in the upwind direction,
\footnote{An independent sorting of local interstellar velocity
  absorption components into clouds is provided by \citet{RLIV:2008}
  and we use that cloud-naming scheme here.}  the shocks that form at
the supersonic interfaces will drive compressional disturbances
into the LIC that perturb the density and velocity of the gas particles.

A supersonic interface would form between the LIC and GC if they are
in contact.  The angle between the LIC and GC LSR velocity vectors is
$17.2^\circ$, and the relative velocity is 14.6 \kms.  The G-cloud has
a temperature of $5500 \pm 400$ K \citep{RLIV:2008}, corresponding to
an acoustic velocity of $\sim 7.7$ \kms, indicating a shock strength
of Mach$\sim 1.9$ between the clouds.\footnote{Using instead the
  heliocentric velocities for the LIC and GC in the upwind direction,
  and now-obsolete physical parameters for both clouds,
  \citet{GrzedzielskiLallement:1996} have predicted that a supersonic
  interface is present between the LIC and the G-cloud, with the LIC
  corresponding to the post-shock gas because it is warmer than the
  G-cloud.  \citet{FrischYork:1991} also modeled the local shock
  structure.}  \nocite{FrischSchwadron:2014}

A supersonic interface would also form between the LIC and the BC if
they are in contact.  These clouds have LSR velocities of $15.3 \pm
2.5$ \kms\ and $7.1 \pm 2.3$ \kms\ respectively, and an angle between
the two vectors of 49\deeg\ (cloud LSR velocities are from Frisch \&
Schwadron 2014).  The LIC sound velocity is 8.6 \kms, for a 7000 K
perfect gas with mean molecular weight 1.29.  The $\sim 11.9$
\kms\ relative velocity indicates a supersonic interface of Mach
number $\sim 1.4$ between the two clouds if they are in direct
contact.  The LIC is warmer than the BC, which has a temperature
$3000^{+2000}_{-1000}$ K \citep{Hebrard:1999}.  A simple temperature
gradient between the three clouds does not exist.  In the presence of
the interstellar magnetic field, the shock transitions will be
modified by the unknown angle between the magnetic field and the
normal to the cloud surface.

\section{Refitting the historical data on the interstellar wind longitude} \label{sec:refit}

A new statistical interpretation of the historical data (Table 1 in
the online supplementary material, OSM, of Paper I) is required by a
typographical error in the original NOZOMI publication
\citep{Nakagawa:2008nozomi}.  We first perform new fits to the same
set of data used in the Paper I using the corrected uncertainty for
the \HeI\ flow longitude determined from the NOZOMI data \citep[][\S
  \ref{sec:584n}]{Nakagawa:2014erratumnozomi}, and next for an
alternate selection of the fitted data.

\begin{figure}
\plottwo{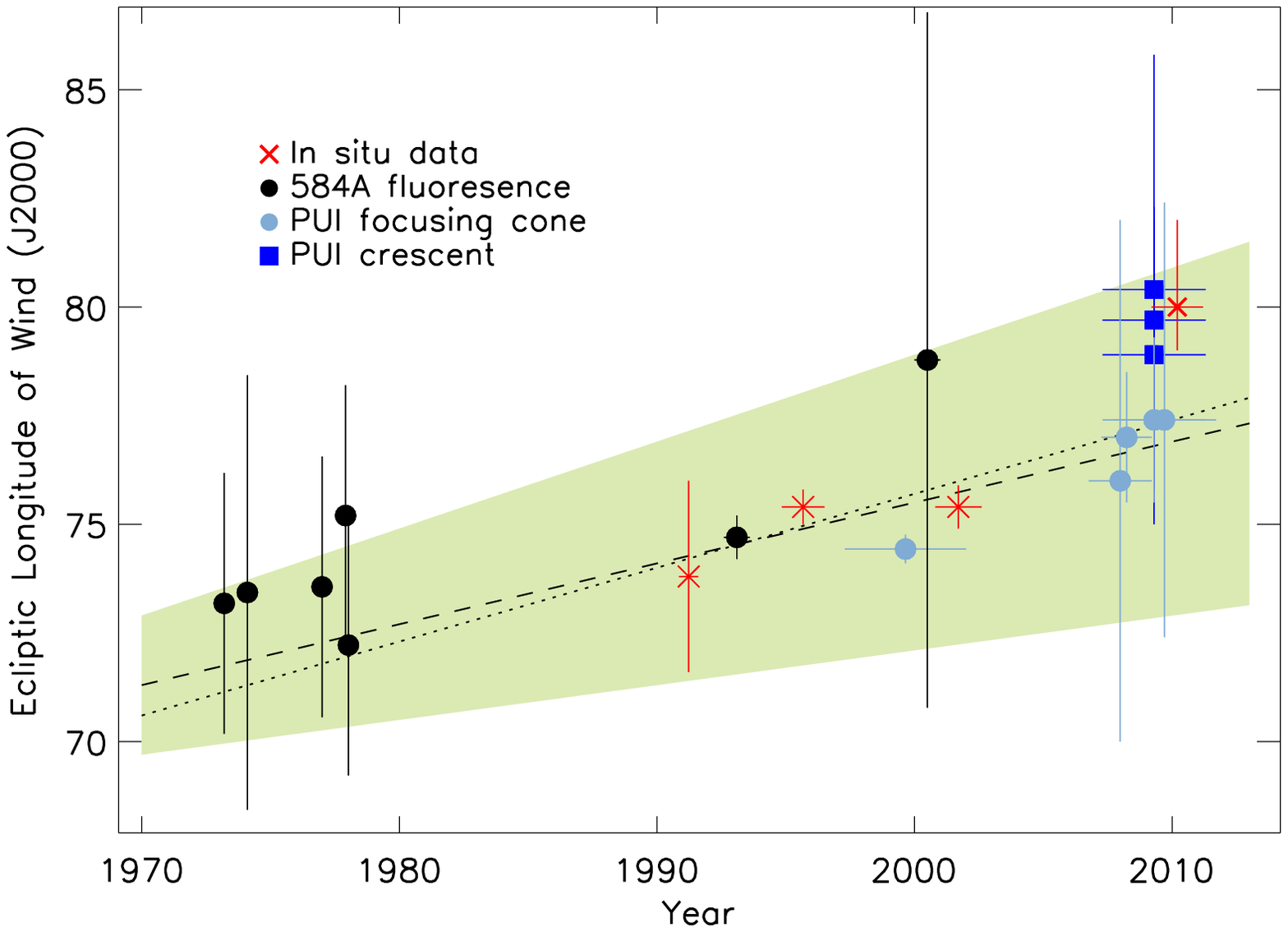}{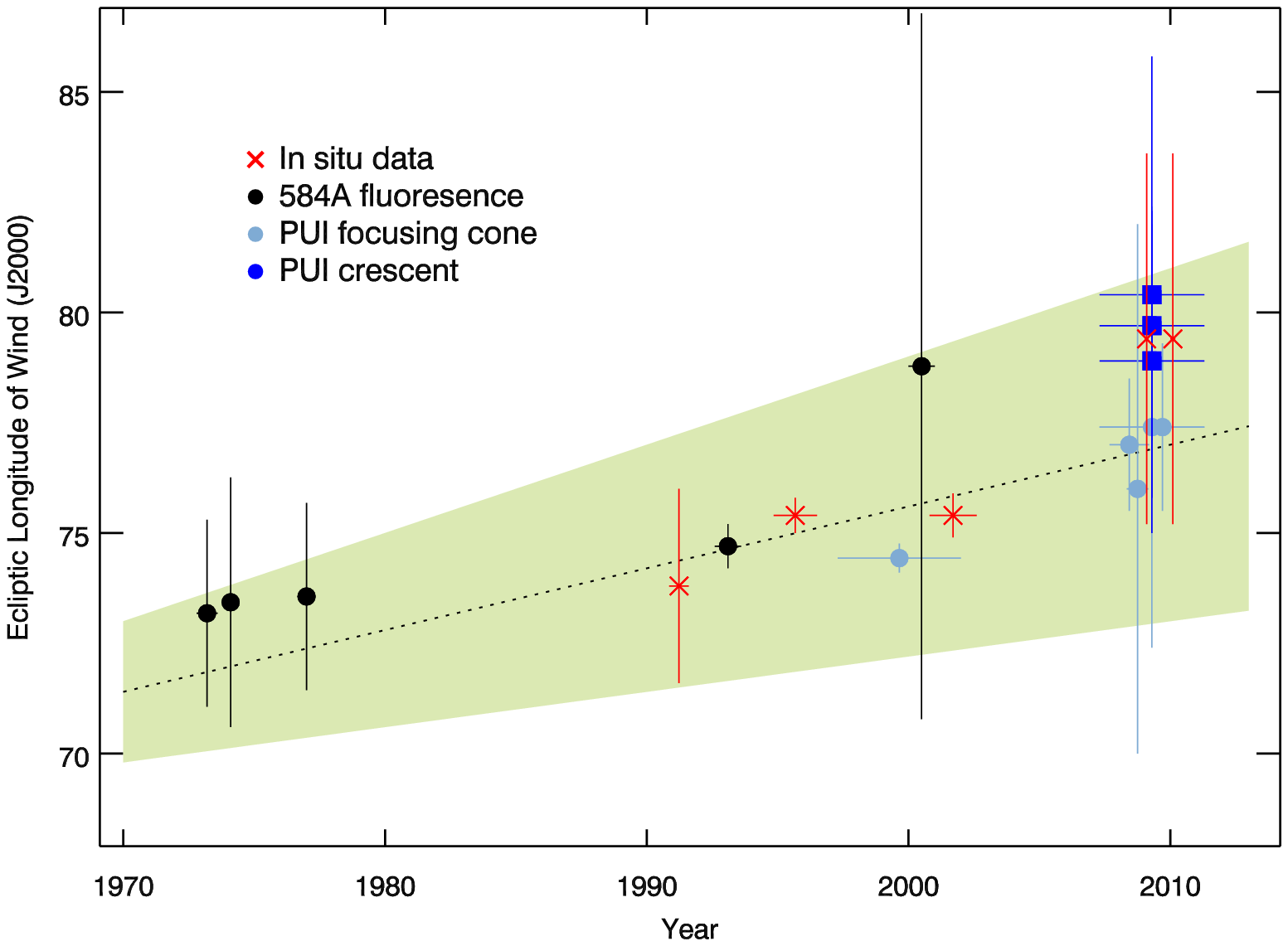}
\caption{Left: The fit using the correct NOZOMI uncertainties is shown
  by the dashed line, and the dotted line shows the original (Paper I)
  fit with the uncorrected NOZOMI uncertainties.  The shaded region
  shows the uncertainties for the new fit.  The uncertainties on
    the IBEX data point (red dot in upper right) are constrained by
    the IBEX parameter range and the LIC temperature toward Sirius
    based on \FeII, \MgII, and \CaII\ data (Paper I, OSM-S4).  Right:
  Same fit as on the left, except that the Prognoz 6 data are omitted,
  the uncertainties for the STP72-1, Mariner 10, and SOLRAD11B data
  are revised to include only the longitude uncertainty, and seasons
  2009 and 2010 of the IBEX data are listed as independent data
  points. The IBEX longitude in the right hand figure does not depend
  on any constraints that are based on temperatures obtained from LIC
  absorption lines.  These fits are explained further in Section \ref{sec:refit}.
  }\label{fig:refit}
\end{figure}

The results of the refitting are presented in Table \ref{tab:refit}.
The first result, heading A, shows the original fit from Paper I.  The
updated fit using the corrected NOZOMI uncertainties is listed under
heading B, fits 3 and 4, of Table \ref{tab:refit}, and plotted in
Figure \ref{fig:refit}.  The slope of the revised fit is $0.14 \pm
0.06$, which implies a directional change of $\delta \lambda =
5.6^\circ \pm 2.4^\circ $ over the past 40 years (instead of the
$6.8^\circ \pm 2.4^\circ$ in Paper I).  The reduced $\chi^2$ of the
fit is 0.7, and the p-value of 0.82 is larger than the 0.05 value
required to pass the p-test
\footnote{As pointed out in S5 of the OSM in Paper I, the probability
  of taking a result $\chi^2$ more extreme than the observed value of
  $\chi^2_{obs}$ is given by the p-value that equals the minimum
  between the two probabilities, $P(0\le \chi^2 \le
  \chi^2_\mathrm{obs}$) and P($\chi^2_\mathrm{obs} \le \chi^2 \le
  \infty$).  A null hypothesis associated with a p-value smaller than
  the significance level of 0.05 is typically rejected.  Additional
  information about the statistical techniques used in the fits can be
  found \citet{Livadiotis:2014}, \citet{Livadiotis:2007},
  \citet{Deming:1964}, \citet{Lybanon:1984}} so that this result is
highly likely.  The small value for $\chi^2$ suggests that the
uncertainties may be overestimated by about $\sim 20$\%.  The
corresponding fit for the assumption of a constant flow longitude
gives $ \lambda = 75.0 \pm 0.3$ (fit 4), which has a p-value of 0.014
and is now a statistically acceptable fit, but with a lower likehood
than a fit with a variable flow direction.  This new fit to the data
using the corrected NOZOMI uncertainties should replace the fit that
is presented in Paper I.

In order to test the robustness of the fit to these data, we have
performed a second fit with four additional modifications made to the
input data set:
\begin{enumerate}
\item
The uncertainties assigned to the STP 72-1, Mariner 10, and SOLRAD11B
flow directions in Paper I represented the combined longitude and
latitude uncertainties reported in the original publications (\S
\ref{sec:584}).  The new fit uses only the longitude uncertainty.  It
is assumed that the uncertainties for longitude and latitude are
normally distributed, so that the uncertainty used in Table 1 of Paper
1 can be reduced by a factor of 1.414.
\item
The IBEX and Ulysses data are treated equivalently in the sense that
the two independent data sets corresponding to the 2009 and 2010
seasons are included separately, rather than combined as in Paper I.
The longitudes for the individual seasons from \citet{Bzowski:2012isn}
are used.  

\item The uncertainties for the IBEX longitudes are based on the IBEX
  data alone, without recourse to astronomical data on the LIC.  The
  IBEX temperature uncertainty range of 4400-8200 K (\S
  \ref{sec:ibexseason}) can be used to set longitude uncertainties of
  $\pm 4.2^\circ$ \citep[Fig. 1 of][]{McComas:2012bow}.  In Paper I a
  LIC temperature based on astronomical data was used, however the
  astronomical data reveal large variations in the LIC temperature
  toward different stars (Appendix \ref{app:lic}).

\item The Prognoz 6 data are omitted from this fit because both the
  geometric fit and the fit obtained from modeling the data are no
  longer considered valid by coauthors of the original papers (Section
  \ref{sec:584p}).
\end{enumerate}

Three fits were performed on this adjusted data set, a linear fit, a
fit assuming an constant flow longitude, and a parabolic fit (heading
C, fits 5, 6 and 7).  The goal of the fits is to test whether a
temporal variation in the flow direction over time is statistically
more likely than a constant flow direction over time.  The linear fits
listed in Table \ref{tab:refit} are highly likely and always more
likely than a fit with a constant longitude over time.  The low
$\chi^2$ value of the parabolic fit suggests that the limited set of
available data does not justify a second-order fit.  Note that the
finding that a linear fit to the data is statistically more likely
than a constant or parabolic fit is not equivalent to a statement that
a change in longitude must have occurred in a linear fashion.

The linear fit is statistically the most likely fit and gives a
temporal variation of $ 5.6^\circ \pm 2.4^\circ$ over the past 40
years (the fit is shown in Figure \ref{fig:refit}, right).  This
result is close to the result in Paper I.  However while a constant
longitude could be rejected in Paper I, the increase in the NOZOMI
uncertainties makes a constant longitude over time acceptable but less
likely than a variation of the longitude over time \citet{Livadiotis:2014}.

The parabolic fit is included to show that a fit with a higher order
polynomial yields a result that is less satisfactory than a fit with a
lower order polynomial.  As the order of the polynomial in the fit
increases, the \chisq\ should in principle decrease until it is $<1$
or it is \chisq$\sim 1$.  The parabolic fit under heading C gives
\chisq$\sim 0.6$, which tells us that the best statistical description
of the temporal variation in the flow direction is given by the linear
fit.

New reductions and modeling of the Ulysses data have appeared since
the publication of Paper I
\citep{Bzowski:2014ulysses,Wood:2014ulysses,KatushkinaWood:2014ulysses}.
These results find a \HeI\ flow longitude for the Ulysses data
collected during the 2007 polar pass that is close to the value of
\citet{Witte:2004}, but with a somewhat higher temperature
\citep[$7500^{+1500}_{-2000}$ K,][]{Bzowski:2014ulysses}.  The
differences between the \HeI\ flow vector found from the IBEX
2009--2010 seasons of data
\citep{Moebius:2012isn,Bzowski:2012isn,McComas:2012bow}, and the
Ulysses vector, are not yet explained.  For the latest seasons of IBEX
observations \citep[2012--2014,][]{McComas:2014warm,Leonard:2014isn},
the results from spacecraft pointings in the ecliptic suggest a
different portion of the coupled 4-dimensional ``tube'' of possible
solutions presented by \citet{McComas:2012bow} with similar
\HeI\ velocity vectors to Ulysses but requiring a significantly warmer
cloud; other new IBEX pointing directions, at different angles to the
ecliptic plane, are not consistent with the Ulysses data for reasons
not fully understood.  Comparisons of Ulysses and IBEX results are a
study-in-progress and we do not include results presented after 2013
in the refitting of the data in this paper.

\section{Data used for detecting temporal variations in the neutral helium wind} \label{sec:data}

LB14 have argued that most of the data incorporated into the
meta-analysis of Paper I were either incorrect or were used
incorrectly.  These data, all of which were taken from the refereed
scientific literature, fall into three groups: remote measurements of
the backscattered solar 584\AA\ emission from interstellar \HeI\ in
the inner heliosphere
\citep{WellerMeier:1974,Meier:1977,BroadfootKumar:1978,Ajello:1978,Ajello:1979,WellerMeier:1979,WellerMeier:1981,DalaudierBertaux:1984,FlynnVallerga:1998,Vallerga:2004,Lallement:2004prognoz,Nakagawa:2008nozomi,Nakagawa:2014erratumnozomi};
in situ measurements of the pickup-up ions that are sampled by
spacecraft traversing the gravitational focusing cone and upwind
crescent \citep{Gloeckler:2004,Drews:2012,Gershman:2013}, and in situ
measurements of interstellar \HeI\ atoms
\citep{Witte:2004,Moebius:2012isn,Bzowski:2012isn,McComas:2012bow,Bzowski:2014ulysses}.

The main LB14 arguments are summarized in Table \ref{tab:diff}, along
with our comments on those arguments.  Most of the arguments are
speculative. The LB14 hypothesis about an underestimated IBEX-Lo
instrument response function is wrong (\S \ref{sec:ibexirf}, \S
\ref{app:ibex}).  The better-supported argument that the STEREO pickup
ion data only provide upper limits because of the azimuthal transport
of pickup ions inside of 1 AU has since been shown to be incorrect by
the 0.1 AU parallel mean-free-path of newborn pickup ions found by
\citet{Gershman:2014messenger} from MESSENGER data (\S
\ref{sec:stereo}).

\subsection{Evaluation of \HeI\ flow longitude from IBEX measurements} \label{sec:ibextrajectory}

\subsubsection{Functional dependencies of \HeI\ flow direction on peak neutral fluxes } \label{sec:ibexfct}

\begin{figure}[t!]
\plotone{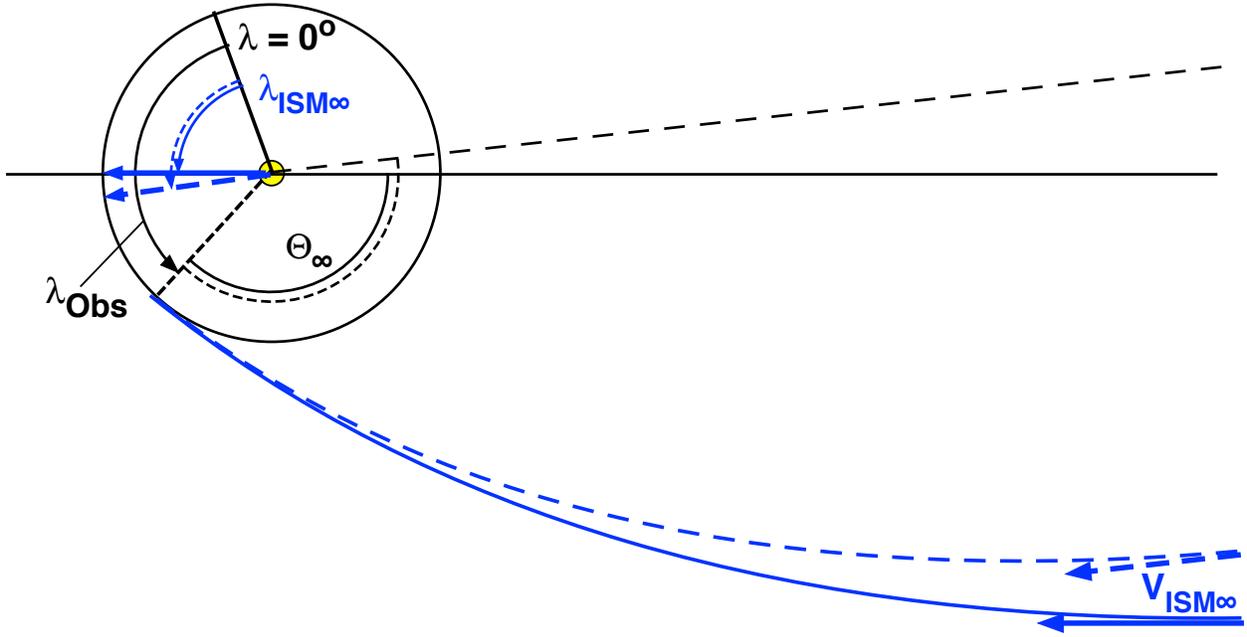}
\caption{ Hyperbolic trajectory geometry in the plane of the ISN bulk
  flow. The location of the perihelion (or flow tangential to the
  Earth's orbit) is degenerate in all combinations \velinf(\laminf)
  that satisfy equation (1).  The solid and the dashed blue line
  indicate two sample trajectories from infinity to their perihelion
  at $\theta=0^\circ$ that satisfy the relation, but arrive from
  different inflow longitudes \laminf\ as indicated by the solid and
  dashed black lines.  $\lambda_\mathrm{Obs}$ is the longitude of the
  observations, \theinf\ is true anomaly of the inflow trajectory
  (blue line) from infinity to perihelion at $r_\mathrm{E}=1$ AU. The
  dashed blue line shows the trajectory of an atom with a different
  \theinf-\velinf\ combination that satisfies equation (1).
}\label{fig:hyper}.
\end{figure}

The trajectories of neutral interstellar helium atoms through the
heliosphere depend on the four parameters describe the interstellar
\HeI\ gas at ``infinity'' beyond the influence of the heliosphere, the
gas flow longitude, \laminf, latitude \betinf, speed \velinf, and
temperature \teminf.  In the parameter space measured by IBEX-Lo,
these parameters are interdependent with coupled uncertainties.  As
discussed by \citet{Moebius:2012isn}, \citet{Lee:2012isn} and
\citet{Moebius:2014icns}, since IBEX is nearly a Sun-pointed spinner,
the determination of the interstellar neutrals (ISN) inflow vector is
tied to the IBEX-Lo observation of the flux maximum that is in a
direction that is perpendicular to the Earth-Sun line.  Then, as shown
in Figure \ref{fig:hyper}, the observed flux maximum is directly
related to the bulk flow vector at 1 AU, in the plane of the ISN bulk
flow trajectory.  Here \velinf\ and \laminf\ are uniquely connected by
the hyperbolic trajectory equation (eqn. \ref{eqn:1}) and the relation
between the true anomaly \theinf\ (the angle between the observed
velocity and the flow vector at infinity) of the trajectory, the
observer location when obtaining the peak \lampeak, and inflow
direction \laminf, so that \theinf=\laminf +$180^\circ$-\lampeak\ is:

\begin{equation}\label{eqn:1}
V_\mathrm{ISM,\infty} =       
\left[\frac{GM_\mathrm{S}}{r_\mathrm{E}} \left(\frac{-1}{\cos(\theta_{\infty})} -1\right)\right]^{1/2} 
\end{equation}

Without further information, this constitutes a functional
relationship of \velinf(\laminf).  The location of the flow maximum
observed by IBEX along the Earth orbit \lampeak\ was determined to be
$ 130.6^\circ \pm 0.7 ^\circ$ in \citep{Moebius:2012isn}, which constrains the
functional relationship in eqn. 1 to within $\le \pm 0.7 ^\circ$ in
\laminf\ and $\le \pm 0.5$ \kms\ in \velinf, although a large range of
values along the function satisfy the observed location of the
interstellar gas flow maximum at 1 AU nearly as well.

\begin{figure}
\plotone{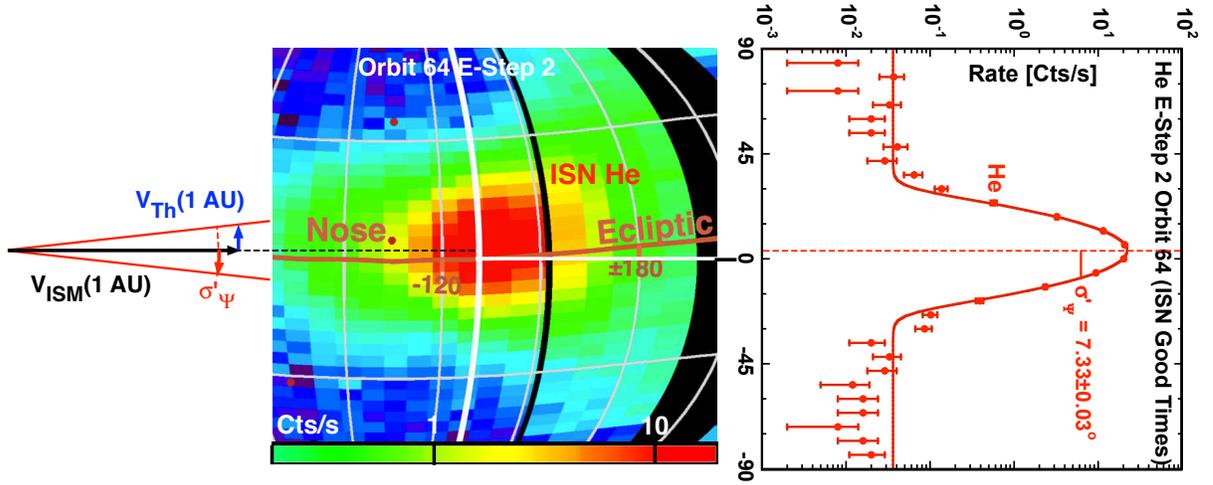}
\caption{Composite of angular distribution measurement as a local Mach
  Cone, whose width (shown on the left) $\sigma^\prime_\psi =
  v_\mathrm{th(1 AU)}/V_\mathrm{ISM(1 AU)}$ is defined by the ratio of
  the local thermal and bulk velocities of the flow.  The angular
  width in latitude $\sigma^\prime_\psi$ is obtained from the cut
  through the maximum of the neutral helium rate distribution in
  ecliptic longitude (along the white arc) shown in the color coded
  map (center panel). The latitudinal rate distribution of that cut
  (right panel) is shown along with a Gaussian fit and statistical
  uncertainties.  }\label{fig:swath}
\end{figure}

Simultaneously, the ISN flow direction in latitude \betinf(\laminf)
and the ISN temperature \teminf(\laminf) are determined in the
\citet{Moebius:2012isn} analysis from the ISN flow peak location in
latitude and its angular width in latitude, $\sigma_\psi$, the
observation of which is shown in Figure \ref{fig:swath}, again varying
with the ISN flow longitude \citep[also see][]{Moebius:2014icns}.  As
indicated in the left panel of Figure \ref{fig:swath}, the observed
angular distribution of the incoming interstellar neutrals constitutes
a local Mach cone of width $v_\mathrm{Th}$/\veloneau, based on the
bulk flow \veloneau\ and the thermal velocity $v_\mathrm{Th}$ of the
distribution at 1 AU. Therefore, the observed angular width fixes the
ratio \teminf$^{1/2}$/V$_{\mathrm{ISM},1 AU}$, and the derived
temperature depends on the derived ISN bulk flow speed \velinf, the
only variable in \veloneau, and thus on \laminf.  Because the
statistical uncertainties of the observed peak location and width of
the ISN flow are only 2\% and 1\%, respectively, the combined
uncertainties of the two relations are mainly determined by the
uncertainties in relation (1) and potential systematic effects. We
will address such effects below, in particular, on the resulting
temperature.

\begin{figure}
\plotone{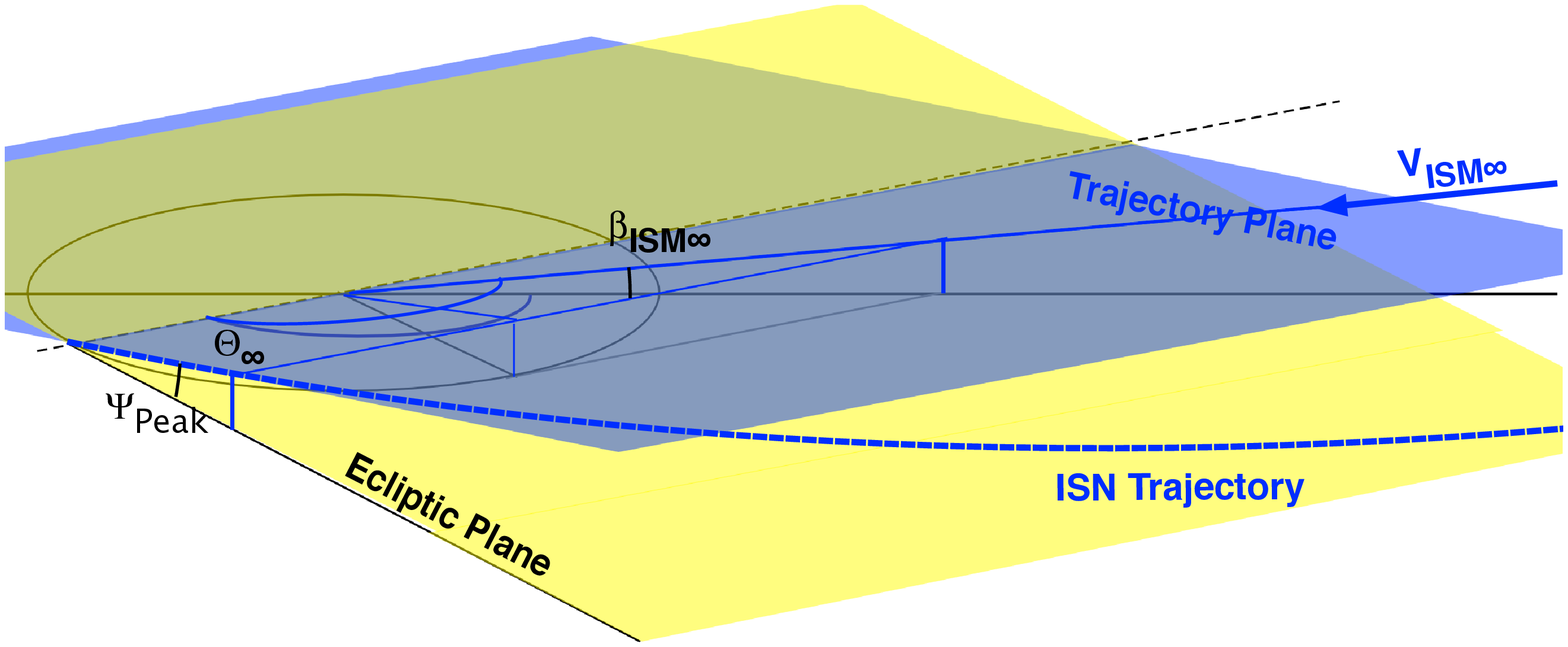}
\caption{Schematic view of the interstellar gas flow in its trajectory
  plane (blue), which is inclined to the ecliptic (yellow) at the
  latitude inflow angle \betinf.  $\psi_\mathrm{Peak}$ is the latitude
  of the flow in the solar rest frame, \theinf\ is the perihelion
  angle with respect to the interstellar flow at infinity, and
  \betinf\ is the latitude of the interstellar flow at infinity.  The
  x,y,z (and primed) coordinates define the cartesian coordinate
  systems.  }\label{fig:plane}
\end{figure}

As shown in Figure \ref{fig:plane}, the latitude $\psi_\mathrm{Peak}$
of the flow at 1 AU in the solar rest frame is connected to the ISN
flow latitude \betinf\ at infinity through the true anomaly of the
bulk flow as
\begin{equation}\label{eqn:2}
\tan(\psi_{Peak}) = \frac{\tan(\beta_{ISM,
    \infty})}{|\sin(\theta_{\infty})|} =
\frac{\tan(\beta_{\mathrm{ISM,\infty}})}
     {|\sin(\lambda_{\mathrm{ISM,\infty}} + 180^\circ -
       \lambda_\mathrm{Peak})|}
\end{equation}
and thus again to the ISN inflow longitude \laminf. These relations
constitute a degeneracy in the determination of \velinf, \betinf, and
\teminf\ as a function of \laminf. However, relation (2) also provides
a tool to break the degeneracy using IBEX latitude distributions
combined for different positions of IBEX as the observer varies in
longitude \lamobs. Then, \lamobs\ replaces \lampeak\ in equation (2)
as a variable and the resulting relation $\psi$(\laminf) depends
distinctly on \laminf.  The comparison of ISN flow distributions
observed over entire 2009--2010 observation seasons with the predicted
relation (2) as a function of \laminf\ and respective
$\chi^2$-minimization in \citet{Moebius:2012isn} resulted in the
reported optimum inflow longitude \laminf=$79.0^\circ\ (+3.0^\circ,
-3.5^\circ)$ with a similar result in Bzowski et al. (2012).
\citet{Bzowski:2012isn} used a $\chi^2$ analysis to find an
uncertainty range for the interstellar flow longitude of $ \sim \pm
4\degr$ \citep[blue contour in Fig. 22 of][]{Bzowski:2012isn}.  These
uncertainties were the basis for the bounding values of the resulting
parameter range from the IBEX observations used in McComas et
al. (2012).

In contrast, Lallement \& Bertaux (2014, in top box of Fig. 2)
maintain that the variation of the observed fluxes as a function of
observer longitude \lamobs, with a substantially wider maximum than
observed with Ulysses GAS (Witte 2004), may have led to the different
inflow longitude, which may be affected by instrumental influences. To
support their claim, they compare the observed IBEX count rate
distributions with simulated distributions for the Witte (2004)
parameter set produced by Bzowski et al. (2012) from the best-fitting
$\chi^2$ analysis.  However, the difference in the distributions
between these two sets is totally dominated by the larger angular
widths of the count rate sequences measured by IBEX, which leads to
the larger derived temperature, and not by unaccounted-for data
losses.  As shown by Lee et al. (2012), the widths of the distribution
in latitude and longitude are tightly coupled and represent the ISN
temperature.

\begin{figure}
\plotone{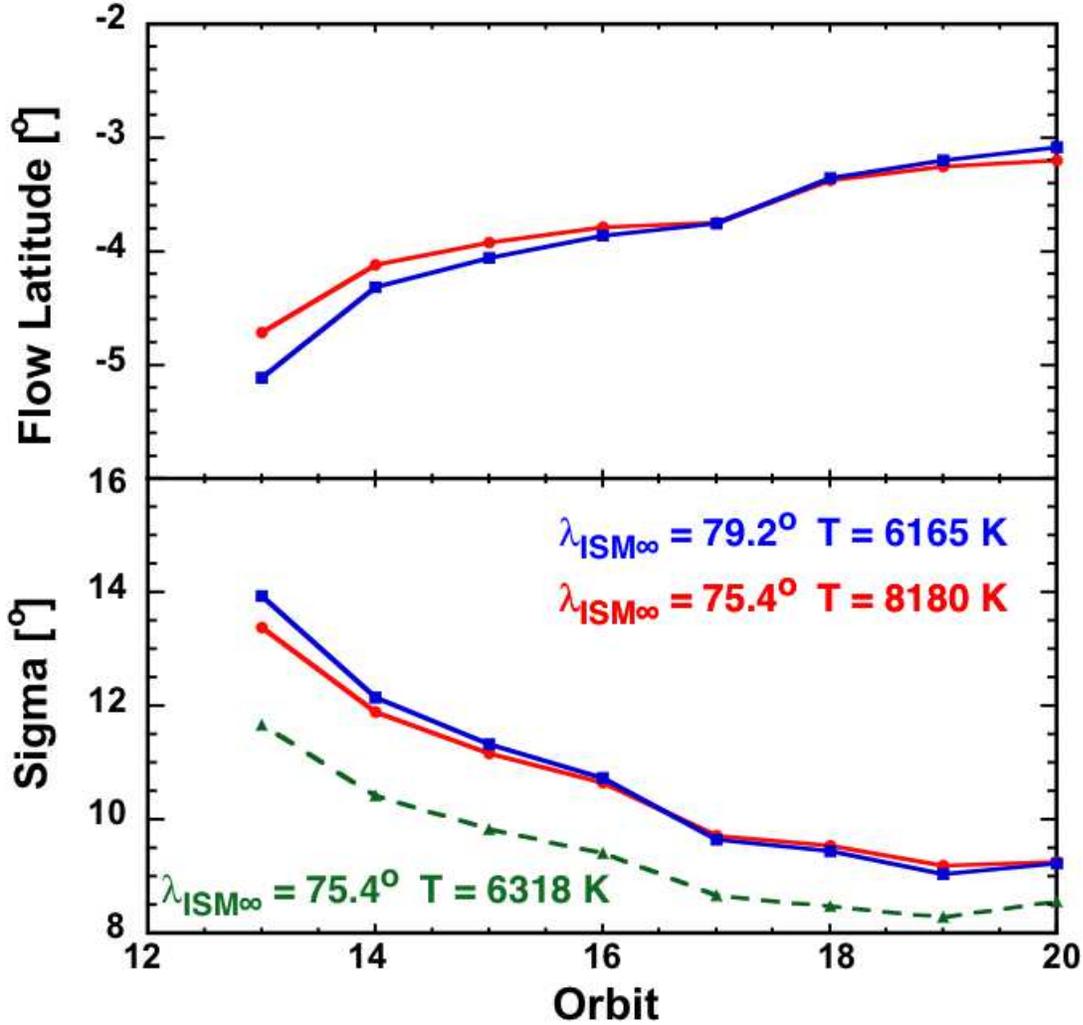}
\caption{Peak location in latitude (top) and width (bottom) of ISN
  flow distributions for the IBEX orbits during the 2009 ISN flow
  season, simulated for the actual observation periods with two ISN
  flow parameter sets along the center of the IBEX tube in parameter
  space.  The inflow longitude of Witte (2004) is shown in red and the
  one of Bzowski et al. (2012) and \citet{Moebius:2012isn} in blue.
  The latitudinal width of the distribution, $\sigma$, is also shown
  in green for the temperature found by Witte (2004).  These
  simulations are based on the measured properties of the IBEX-Lo
  instrument.  A temperature of 8180 K would be required for the Witte
  (2004) flow vector to match the latitudinal width measured by IBEX.
}\label{fig:peak}
\end{figure}

To illustrate the basis for the inflow direction in the IBEX analysis,
Figure \ref{fig:peak} shows the peak location and width of the ISN
flow distributions in latitude as simulated for the observation
periods during the 2009 ISN season orbits. Shown are the results for
two ISN flow parameters sets that lie on the center line of the
parameter tube \citep[e.g. Fig. 1 in][]{McComas:2012bow} as defined in
equations (1) and (2), using the inflow longitude \laminf\ of Witte
2004 (red) and of Bzowski et al. and
\citet[][blue]{Moebius:2012isn}. For comparison the width is shown as
obtained with the temperature found by Witte (2004), which stands out
as much narrower than that from IBEX. However, the visible differences
between the two parameter sets along the IBEX tube are, first, a
steeper variation of the peak location as a function of orbit for
\laminf\ as favored by Bzowski et al. (2012) than for the value of
Witte (2004) and, second, a somewhat stronger increase in width
towards the early orbits. To arrive at a value for \laminf, solely the
first effect has been used by \citet{Moebius:2012isn}.  In relation to
\laminf\, the $\chi^2$-minimization in the comparison between observed
and simulated flow distributions used by Bzowski et al. (2012, see
Fig. 22) is sensitive to both effects as illustrated in Figure
\ref{fig:peak}.

This leaves only the argument of LB14 that the width of the angular
distribution, which now solely influences the derived temperature,
could be brought into agreement with the Ulysses value by considering
the influence of a strong and unaccounted-for instrumental effect in
the IBEX data reduction.  As derived from extensive tests with the
actual IBEX data system hardware simulator and Monte Carlo simulations
of realistic data throughput of a wide range of particle rates, the
actual effect of data suppression due to limited transmission
bandwidth on the width of the observed distributions is $<2$\% and
thus is negligible (see Appendix \ref{app:ibex}, including Figure \ref{fig:rate}).

LB14 make the point that the degeneracy between the longitude and
velocity of the \HeI\ distribution function allows the longitude of
the peak count rates of IBEX data to also correspond to the
expectations for the Ulysses flow vector, providing there is an
additional suppression of count rates (detector ``dead-time'') that
was neglected in the IBEX data flow.\footnote{More specifically, they
  state that ``We have examined the consequences of dead-time counting
  effects, and conclude that their inclusion at a realistic level is
  sufficient to reconcile the [IBEX 2009--2010] data with the
  \emph{old} [Witte Ulysses] parameters, calling for further
  investigations.''}  In support of their hypothesis that the helium
vector of \citet{Witte:2004} is correct, LB14 took the orbit-by-orbit
IBEX data, and the simulations where
\citet[][Fig. 18]{Bzowski:2012isn} compared the IBEX data with the
Ulysses flow parameters.  Those rates and data simulations were
replotted for two cases (Fig. 2 of LB14).  One line on the plot showed
the results of the \citet{Bzowski:2012isn} analysis, while the other
line plotted an adjusted count-rate based on the LB14 hypothesized 7
ms dead-time correction for the IBEX-LO instrument.  They argued that
the Witte (2004) flow direction agreed with the IBEX count rates for
their hypothesized dead-time correction.\footnote{LB14 state: "...it
  is premature to use the new longitude and velocity [of IBEX] before
  the question of the dead-time correction is fully settled."} This
argument to reject the IBEX analysis because of a hypothesized 7 ms
dead-time correction is not valid.

The effect that influenced the IBEX data flow is a known finite data
processing time, not a sensor dead-time, and the types of events
mostly responsible for this effect were unanticipated electrons, which
do not alter the shape of the observed interstellar flow distribution
significantly.  Also, these are excluded by a new instrument mode that
started mid-2012.  This operational modification has eliminated the
count rate reduction reduction effect almost entirely.  Comparison
between the old and new operational modes has demonstrated that the
finite data processing time does not significantly affect the count
rate distributions used to reach the conclusions reported in
\citet{Moebius:2012isn}, \citet{Bzowski:2012isn}, and
\citet[][Appendix \ref{app:ibex}]{McComas:2012bow}. In particular, the
data processing time did not influence the deduced He temperature,
which LB14 tried to reconcile by exploring this effect to an extent
that was already excluded by \citet{Moebius:2012isn}.  The instrument
response function of IBEX-LO does not contribute in any significant
way to the uncertainties of the interstellar \HeI\ flow direction
derived from the IBEX data.  This is discussed in more detail in
Appendix \ref{app:ibex}.

In summary, the inflow longitude parameter hinges mostly on the
evolution of the peak location in latitude as a function of observer
longitude in the IBEX observations. While the temperature in
\citet{Moebius:2012isn} is obtained by fitting the height and width of
the count rate distributions, these observables were ignored in LB14.
The temperature derived from the IBEX observations, for the case where
one assumes the inflow velocity of the Ulysses results (Witte, 2004),
is 30\% higher than the temperature reported from Ulysses GAS
observations of \HeI.  Systematic instrumental effects larger than 2\%
in the IBEX observations, such as an erroneous instrument response
function, can be ruled out as the reason for the difference between
the Ulysses and IBEX \HeI\ flow directions ( also see Appendix
\ref{app:ibex}).

\subsubsection{IBEX results for 2009--2010 seasons of data} \label{sec:ibexseason}

IBEX-Lo directly samples the trajectories of interstellar He atoms at
1 AU after they have survived photoionization, electron-impact
ionization, and charge-exchange with the Solar wind to reach the 1 AU
orbit of IBEX
\citep{McComas:2009sci,Fuselieretal:2009ssr,Moebius:2009sci,Fuselier:2009sci,Moebius:2012isn,Bzowski:2012isn,Moebius:2014icns}.
IBEX-Lo collects interstellar helium atoms over several orbits during
late winter and early spring where the atoms are near perihelion of
their trajectories \citep{Lee:2012isn}.  In the earliest orbits,
IBEX-Lo samples the ``warm breeze'' that consists of secondary
\HeI\ atoms that are deflected beyond the heliopause
\citep{KubiakBzowski:2014HeO}.  After that the direct interstellar
\HeI\ wind dominates the count rate.

The interstellar helium wind direction was obtained from these
measurements using two independent analysis methods.
\citet{Moebius:2012isn} used an analytical approach to fit the
evolution of the helium count rates across the orbits where helium is
observed, where the outer boundary of the computations was at
infinity.  \citet{Bzowski:2012isn} compared simulations of the helium
distribution obtained from the Warsaw Test Particle model with the
IBEX data, and performed $\chi^2-$fitting between the models and data
to obtain the best helium wind parameters, where the outer boundary of
the computations was at 150 AU, the distance at which the interaction
is already underway.  \citet{McComas:2012bow} resolved the small
discrepancy caused by the different outer boundaries used in the two
computations and showed that these independent analyses of the IBEX-Lo
data yield similar results.

Independent analysis of the 2009 and 2010 seasons of IBEX data yielded
the same flow direction within a large acceptable range.  For the IBEX
flow direction during the 2009 observing season,
\citet{Bzowski:2012isn} found a He flow vector of \laminf=79.2\deeg,
\betinf=--5.06\deeg, \velinf=22.831 \kms, and \teminf=6094 K. For the
2010 season, the \HeI\ flow vector was \laminf=79.2\deeg,
\betinf=-5.12\deeg, \velinf=22.710 \kms, and \teminf$=6254$ K.  These
solutions also provided acceptable $\chi^2$ values for mutual fits of
the four interdependent parameters \citep[Fig. 22 in
][]{Bzowski:2012isn}. The acceptable range for the longitude is
$75.2^\circ - 83.6^\circ$, corresponding to temperatures of 4400--8200
K, with each flow longitude coupled to a tightly constrained range of
the other parameters because the uncertainties are correlated.  In
contrast, when the Ulysses helium flow longitude \citep{Witte:2004}
was used as a constraint to the IBEX longitude data, the best reduced
$\chi^2$ fit was an order-of-magnitude larger \citep[Fig. 17 in
][]{Bzowski:2012isn}.  Independently, and with the analytic method
described by Lee et al. (2012), \citet{Moebius:2012isn} determined a
flow vector for the combined 2009-2010 seasons of
\laminf$=79.0^{+3.0}_{-3.5}$ deg, \betinf$ =-4.9 \pm 0.2$ deg,
\velinf$=23.5^{+3.0}_{-2.0}$ \kms, and \teminf$=6200 ^{+2000}_{-1200}$
K.  The best-fitting direction of the interstellar helium wind using
two different analysis techniques are in good agreement with each
other.

The large range of possible longitudes that is obtained when the
correlated uncertainties of the 4-dimensional parameter are included
can be constrained further if independent information on the LIC
temperature is available.  In Paper I we used the gas temperature that
is obtained from astronomical observations of the spectra of
refractory interstellar elements toward Sirius (see \S \ref{app:lic}).

One outcome of the Bzowski et al. (2014) reanalysis of the Ulysses
data set is that the ``suspicious'' coincidence remarked upon by LB14,
that the latitude of the \HeI\ flow did not vary between between the
IBEX and Ulysses \HeI\ measurements, may not be so settled after all
with the new reduction and modeling of the Ulysses data.
\citet{Bzowski:2014ulysses} found that the best-fit Ulysses upwind
direction of the flow latitude is one degree north of the center of
the IBEX range, at \betinf$=6^\circ \pm 1.0^\circ$ versus
\betinf$=4.98^\circ \pm 1.21^\circ$ for IBEX \citep{McComas:2012bow}.

\subsubsection{IBEX instrumental response function } \label{sec:ibexirf}

In order to support their perspective that the interstellar wind
direction through the heliosphere is invariant with time, LB14
speculated that the instrumental response function used to analyze the
IBEX-Lo data is incorrect, and argued that the true ``dead-time''
correction is 7 ms, which is larger than the value of the instrument
response function underlying the IBEX data analysis of the 2009--2010
observing seasons.  An explanation of the incorrectness of this
speculation is given in Appendix \ref{app:ibex}, where the details of
the IBEX instrumental function are discussed.

In addition, the ISN flow direction is driven by the variation of the
peak in latitude with the ecliptic longitude of IBEX, which is not
affected by data flow restrictions.  IBEX is a Sun-pointed spinner
that measures vertical swaths of the sky with each spin
\citep{McComas:2009sci}.  The \HeI\ particle trajectories are measured
over both longitude and latitude, and that are insensitive to the
speculations of LB14 regarding the instrumental response parameter
(Appendix \ref{sec:ibextrajectory}).  The best-fitting interstellar
neutral (ISN) flow direction found from IBEX observations 
(\S \ref{sec:ibexseason}) is not dominated by the longitudinal height and
width of the observed flux distribution.

LB14 have also claimed that the second independent analysis of IBEX
data by \citet{Bzowski:2012isn} omits ``dead-time'' corrections.  This
argument is invalid because the Bzowski et al. results are based on
data that used an instrument response function that had been tested
and validated.

\subsection{STEREO measurements of pickup ions} \label{sec:stereo}

Measurements of He, Ne and O pickup ions (PUI) by the STEREO PLASTIC
instrument between 2007 and 2011 are reported by \citet[][hereafter
  D12]{Drews:2012}.  Neutral interstellar He, Ne and O atoms penetrate
to the inner heliosphere where they are ionized by photoionization,
electron-impact ionization, or charge-exchange with the solar wind
\citep{Bzowski:2013HeNeO}.
\footnote{ Photoionization models of the LIC predict neutral fractions
  at the heliosphere for hydrogen, helium, neon, and oxygen of $\sim
  0.78$, $\sim 0.61$, $\sim 0.20$, and $\sim 0.80$ respectively
  \citep{SlavinFrisch:2008}.  Up to 60\%, 30\%, and 4\% of these
  neutral interstellar He, Ne and O atoms, respectively, survive to
  the Earth's orbit before being ionized by a mix of electron-impact
  ionization, charge-exchange with the solar wind, and photoionization
  \citep{Bzowski:2013HeNeO}.  PUI data sample particles created
  between the Sun and the observation point.}  The charged particles
that form inside of 1 AU couple to the solar wind magnetic field and
are convected radially outwards with the solar wind where they can be
observed at 1 AU.  Helium and Ne survive the larger ionization losses
associated with close passage to the Sun and become gravitationally
focused in the downwind direction.  Oxygen is mainly ionized in the
upwind regions.  Helium, Ne and O pickup ions form a region of
enhanced PUI density with a crescent-like distribution in the upwind
hemisphere (D12).  Drews et al. determined the interstellar wind
longitude independently from the He and Ne focusing cones, and the
upwind He, Ne and O crescents, and obtained five independent
measurements of the ecliptic longitude of the interstellar wind
direction (Table \ref{tab:refit}) that were used in Paper I to
evaluate the statistical likelihood that the interstellar wind
direction had changed over the past forty years.

LB14 conclude that the STEREO pickup ion data are only valid as upper
limits on the interstellar wind longitude, and that the oxygen data
should be ignored entirely.  In more detail they argue that: (1)
Pickup ion transport inside of 1 AU causes deviations of the location
of the focusing cone and crescent of up to 5$^{\circ}$ so that values
for the flow directions obtained from STEREO PUI can only be taken as
upper limits. (2) The requirement of a smoothly varying ionization
rate used in the STEREO analysis is not satisfied.  They claim that
D12 only consider photoionization and not considered ionization by
electron impacts.  (3) The inter-comparison of the data between orbit
groupings are invalid because the data are not statistically
independent.  (4) The statistical simulations of the direction
uncertainties of the STEREO analysis are incorrect.  (5) Oxygen should
not be used to infer the interstellar flow since there is also a
population of secondary oxygen atoms.

\def\lmfp{$\lambda_\parallel$}

\subsubsection{Transport of PUIs in the expanding solar wind}  \label{sec:puitransport}

Pickup ions are created in the turbulent environment inside of 1 AU.
If the mean-free-path for transport of the newly-formed PUIs is large,
\lmfp$\sim 1 $ AU, PUI ions are transported parallel to the magenetic
field lines so that their spatial distribution is no longer a valid
marker of the ISN wind direction \citep{Chalov:2014}.
\footnote{\citet{Chalov:2014} has investigated the influence of transport
effects on the helium focusing cone position for the unusual solar
minimum conditions between 2007 and 2009, an interval that is included
in the STEREO measurements.  By numerically solving the transport
equation of \HeII\ pickup ions, the possible displacement of the
helium focusing cone was evaluated for different values of the mean
free path of the atom parallel to the magnetic field, $\lambda_\parallel$.
For the most extreme conditions, $\lambda_\parallel = 1 $ AU gave an
offset angle for the observed focusing cone center of 3\deeg, which
would then need to be subtracted from the measured value to get the
true inflowing wind direction. However for $\lambda_\parallel=0.1$ AU
of the MESSENGER data, no transport is found.}
However if the PUIs are formed in a non-turbulent environment,
e.g. \lmfp$\sim 0.1$ AU, the parallel transport is negligible.
A new MESSENGER study has shown that \lmfp$\sim 0.1$ AU in the region
0.1--0.3 AU from the Sun \citep{Gershman:2014messenger}.
These new results invalidate the main argument of LB14 that the
PUI data should not be used for determining the direction of
the interstellar neutral wind.  However, we reach this conclusion
with the understanding that new modeling of the ionization
of ISNs close the Sun is providing new insights into the
stability of the focusing cone and crescent locations that
may affect these conclusions (J. Sokol, private
communication, 2014).

LB14 based their conclusions regarding the PUIs on a study of
\citet{ChalovFahr:2006}, who found that transport effects could
generate a relative azimuthal displacement of PUIs to positive
longitudes of up to five degrees, but that the transport effect was
insignificant for the range of $w = v_{\rm{ion}}/v_{\rm{sw}}>1.4$ in
the SWICS data from ACE because of instrumental properties.  However
the STEREO data used by D12 is for pickup ions with $w>1.5$ so the
Chalov and Fahr (2006) modeling does not apply.

In addition to the new MESSENGER results, three other sets of
measurements of the ISN flow longitude obtained from focusing cone
measurements after 2007 agree with each other.  The longitude for the
\HeI\ ISN flow found from STEREO data collected 2007--2011 is
$77.4^\circ \pm 1.9^\circ$ \citep{Drews:2012}.  It is consistent with
the ACE SWICS and MESSENGER FIPS data collected over the first two
years of this interval.  Also, \citet[][G13]{Gershman:2013} analyzed
ACE SWICS 1 AU data from 2007--2009 and found a longitude $77.0^\circ
\pm 1.5^\circ$ that was consistent with the ISN flow direction,
$76.0^\circ \pm 6.0^\circ$, collected by MESSENGER at 0.3 AU during
three passes through the focusing cone during 2007--2009.  Gershman et
al. (2013) concluded that the similar longitudes found from the 1 AU
SWICS data and the 0.3 AU FIPS data indicated that azimuthal transport
effects have a minimal effect on the \HeI\ flow longitude, and that
electron impact ionization is significant inside of 1 AU.

However, as noted above, transport effects appear to have a minimal
effect on the direction of the He focusing cone.  STEREO
\citep{Drews:2012}, and MESSENGER and ACE \citep{Gershman:2013}
measurements of the helium focusing cone were collected conducted
during similar temporal intervals of 2007--2009, during solar minimum,
and found focusing cone directions in agreement.  The MESSENGER data
were collected at 0.3--0.7 AU while ACE and STEREO data were obtained
near 1 AU, suggesting that transport effects are insignificant.
However the larger uncertainties on the \HeI\ flow longitude obtained
from the MESSENGER measurements, $77.0^\circ \pm 6.0^\circ$, do not
allow possible transport effects to be ruled out conclusively.  The
IBEX 2009 in situ measurements were collected over a similar time
period and agree with the PUI focusing cone longitude
\citep{Moebius:2012isn,Bzowski:2012isn}.  Comparisons between the ACE
PUI results \citep{Gloeckler:2004} and the Ulysses in situ
\HeI\ results \citep{Witte:2004}, collected at similar times but
different radial distances, identify similar helium flow directions.

Although the possible transport of PUIs inside of 1 AU remains an
important question because of large uncertainties on the helium
focusing cone longitude in the STEREO data, and because of the large
electron impact ionization rate that varies with time, available data
suggest that it is not significant.

\subsubsection{Temporal variations in PUI count rates} \label{sec:stereotime}
The geometric properties of the gravitational focusing cone and upwind
crescent vary with both the solar magnetic activity cycle and the
solar rotation.  The PUIs observed at 1 AU are seeded by the
ionization of neutrals inside of 1 AU \citep{Bzowski:2013HeNeO}, and
subsequent outwards convection as part of the solar wind with a
distinct energy distribution \citep{VasyliunasSiscoe:1976}.

In the approach used by D12, both slowly and rapidly varying
influences on the observed cone and crescent structure are included in
the analysis, but due to the different origins of long-term and
short-term modulations they are included in distinctly different
ways. Slowly changing modulations of the pickup ion count rates
include possible variations of the ionization rate induced by solar UV
radiation or by slow decay of instrumental detection efficiency.
Either would result in a systematic shift of the focusing cone and
crescent location in ecliptic longitude.  These shifts can go in
either direction, and they depend purely on the nature of the
modulation.  Temporal variations in the interstellar wind direction
were assumed to be negligible.

\citet{Drews:2012} estimated the influences of the slowly varying
count rate modulation of interstellar pickup ions on the results and
explicitly included those variations in the analysis.  Comparisons
between SWICS PUI data and the solar UV emission over a ten-year
period of time showed that the influence of slowly varying ionization
rates induced by solar UV radiation is negligible.  Influences of a
slowly decreasing detection efficiency induced by aging of the
PLASTIC's microchannel plates were considered explicitly in the STEREO
analysis by estimating the slowly varying detection efficiency decay
with time. Other slowly varying modulations of the observed pickup ion
count rates, e.g. the effect of the eccentricity of the orbit of
STEREO around the Sun, have been thoroughly discussed in D12.

Pickup ion count rate modulation over \emph{short time scales}, on the
other hand, are included implicitly through the statistical approach
used by D12.  The short-term modulations mainly stem from solar wind
density compressions, changes of the solar magnetic field, or solar
wind variabilities such as the electron impact ionization rates (that
LB14 claim are ignored).  As demonstrated in D12, the statistical
approach is especially well-suited to deal with these kinds of
fluctuations.

The influence of electron impact ionization at 1 AU is strongly
coupled to the occurrence of dense solar wind streams
\citep{Bzowski:2013HeNeO} and therefore fluctuates on short time
scales.  LB14 argue that the statistical analysis does not include the
modulation of pickup ion count rates by electron impact
ionization.\footnote{They comment that ``Drews et al. (2012) consider
  only the photoionization, and omit the second source of ionization,
  electron impact.''}  This statement is misleading because the
statistical analysis focuses specifically on using the statistics of
the total data set as the basis for reliably separating out and
characterizing short-term variations such as electron impact
ionizations.

LB14 make contradictory arguments that the analysis of the
STEREO/PLASTIC data influences of electron impact ionization were
ignored, but also that the ISM flow direction derived from Gaussian
fits to numerically co-added orbital counts, without futher evaluation
of the data, is more accurate.  This second method is based on pure
geometry and explicitly ignores all influences on the formation of
pickup ions including those that are caused by variations in electron
impact ionizations, while the D12 statistical analysis takes those
influences into account.

Further details of the analysis of the STEREO PUI data are discussed
in Appendix \ref{app:stereo}, including the separation of temporal and
spatial variations in the data set, the intercomparison between the
four orbits acquired over the years 2007--2011, and additional
statistical tests on the STEREO results.

\subsubsection{Effect of secondary populations} \label{sec:stereosec}
LB14 argue that the neutral interstellar wind (NISW) longitude that is derived from the oxygen
PUIs in the upwind cresent is invalid because of the presence of
secondary oxygen atoms.  These atoms, however, are inconsequential for
measurements of the oxygen PUIs in the upwind crescent.

The populations of secondary interstellar neutral atoms are produced
by interstellar ions that enter the outer heliosheath regions, and are
first offset by the Lorentz force as they migrate through the
interstellar magnetic field draped over the heliopause, and then
become neutralized through charge-exchange or recombination so that
some of them enter the heliosphere.  IBEX may have detected the
secondary populations of He and O
\citep{Moebius:2009sci,Bzowski:2012isn,Moebius:2012isn} but the
interpretation is still not clear \citep{KubiakBzowski:2014HeO}.
These expected secondary populations are significantly slower than the
primary He and O neutrals \citep{Izmodenovetal:1999}, and therefore
have much lower probabilities of surviving in significant numbers to
the inner heliosphere compared to the primary populations
\citep{Bzowski:2013HeNeO}.  Less than 20\% of the interstellar oxygen
is expected to be ionized \citep{SlavinFrisch:2008} in the LIC, so
there are relatively few atoms in the potential parent population for
the secondary O.  For a likely asymmetric heliosphere, secondary
populations of heavy species such as oxygen will be produced away from
the upwind-downwind axis, where their thermal speeds will be less than
the plasma flow speed in the production region, so they are less
likely to reach the 1 AU region.  When the relatively few secondary
oxygen atoms are coupled to the long travel times through the
heliosphere for secondary populations, combined with the rapid
ionization of oxygen due to photoionization and charge-exchange with
solar wind protons \citep[e.g. see][]{Bzowski:2013HeNeO}, secondary
oxygen is highly unlikely to alter the direction of the interstellar
oxygen flow through the inner heliosphere that is inferred from the O
pickup ions.

The crescent locations deduced from He$^{+}$, Ne$^{+}$, and O$^{+}$
pickup ion observations show a good agreement within the uncertainties
given in D12. Because any secondary flow of neutral helium and neon is
at best at insignificant levels in the analysis, influences by a
secondary oxygen flow are most likely insignificant, as pointed out in
D12.  In addition, any effect on the O$^{+}$ crescent from a secondary
component would push the structure to even higher longitudes than the
one reported in D12.  As \citet{Lallement:2005sci} showed based on
Lyman alpha backscatter observations, the combined inflow direction of
primary and secondary hydrogen is shifted towards smaller longitudes,
which would then be expected also for the combined primary and
secondary oxygen population. \citet{Moebius:2009sci} indeed mentioned
a structure in the neutral interstellar O flow that may be consistent
with a secondary component shifted towards the same side as reported
for H by \citet{Lallement:2005sci}.

\subsection{\HeI\ 584\AA\ backscattered emission} \label{sec:584}

The gravitational focusing cone was first measured in 1972 through the
fluorescence of solar HeI\ 584\AA\ emission from interstellar
\HeI\ close to the Sun \citep{WellerMeier:1974}.  Since then, 40 years
of measurements of the interplanetary 584\AA\ fluorescence form a
historical record of the interstellar wind direction
\citep{Ajello:1978,Ajello:1979,WellerMeier:1981,DalaudierBertaux:1984,Vallerga:2004,Nakagawa:2008nozomi}.
The 584\AA\ data are the earliest measurements of the interstellar
wind direction and velocity.

LB14 question the treatment of the \HeI\ 584\AA\ backscattering data
in Paper I for three reasons: (1) They claim that the \HeI\ wind
direction used in Paper I, and obtained from the Mariner 10 data
\citep{Ajello:1978,Ajello:1979}, is not based only on the
\HeI\ 584\AA\ emission.\footnote{They state: ``We note however that
  the direction quoted in this work has been derived from H
  \lya\ observations, namely from the downwind \lya\ minimum, i.e. it
  applies to the neutral hydrogen and not to helium.''}  (2) They
contend that the value for the flow longitude in
\citet{DalaudierBertaux:1984} that is based on Prognoz 6 measurements
of the geometric center of the peak 584\AA\ emission from the
\HeI\ focusing cone, is inaccurate and should not be used.  (3) They
argue that the uncertainties used for the longitude determined from
the NOZOMI 584\AA\ measurements \citep{Nakagawa:2008nozomi} are
incorrect and larger than the value used in Paper I ($\pm 3.4^\circ$,
that was taken from the original NOZOMI publication).  After Paper I
was published, an erratum for the NOZOMI measurement was published
(see below), so that this argument of LB14 argument was correct.  It
is important to note that this increase in the NOZOMI uncertainties
leads to an statistically possible constant flow longitude of
$75.0^\circ \pm 0.3^\circ$ using the same assumptions as in Paper I,
albeit at a lower likehood than a flow longitude that varies linearly
with time (\S \ref{sec:refit}, Table \ref{tab:refit}).

\subsubsection{ Mariner 10 data:} \label{sec:584m}
Overlooking the discussion of the Mariner 10 data in the online
supplementary material to Paper I, LB14 claim that the longitude of
the NISW used for the Mariner 10 data in Paper I is based on both
\HeI\ and \HI\ data.  The resonant \HeI\ 584\AA\ emission feature was
observed during two roll control maneuvers (RCM) of Mariner 10 --
RCM10 on 28 January, 1974 \citep{Ajello:1978}, and RCM3 on 8 December
1973 \citep{Ajello:1979}.  Both the \HeI\ 584\AA\ and
\HI\ \lya\ emissions were observed during RCM10.  From the
simultaneous modeling of the two RCM10 data sets, Ajello (1978) found
the direction of the interstellar gas flow.  The RCM3 observations
were obtained while the spacecraft was located close to the axis of
symmetry of the \HeI\ focusing cone so that the RCM3 data gave unique
and independent results for the direction of the \HeI\ focusing cone.
The RCM3 \HeI\ wind direction agrees with the wind direction in Ajello
(1978) that is based on both \HeI\ and \HI\ fluorescence data.
\citet{Ajello:1979} state that a change in the location of the
interstellar wind axis by more than $5^\circ$ degraded the fits to the
RCM3 584\AA\ measurements.  The $\pm 5^\circ$ uncertainties in Paper I
are generous since Ajello et al. (1979) quote them for the combined
longitude and latitude, while in Paper I we adopted the full
uncertainty for the longitude alone.

\subsubsection{Prognoz 6 data} \label{sec:584p}
The argument of LB14 that the geometric wind direction from the
\HeI\ 584\AA\ backscattered emission data from Prognoz 6 should not be
used is puzzling because Bertaux is a coauthor on the study
\citep{DalaudierBertaux:1984}.  LB14 object to our use in Paper I of
the wind direction found from the geometric center of the
584\AA\ emission pattern, even through it was quoted with
uncertainties by \citet{DalaudierBertaux:1984}.  The Prognoz 6
analysis was later revised in two papers in order to force agreement
with results from other experiments
\citep{Chassefiere:1988prognozv1v2,Lallement:2004prognoz}.  The
meta-analysis in Paper I incorporated unbiased data published with
uncertainties, so we assumed that both of the \HeI\ wind directions
from Prognoz 6 were useful. However, in light of the LB14 comments and
the later revisions of the Prognoz 6 results, we provide a new fit to
the historical data that omits both of the Prognoz 6 data points
(Section \ref{sec:refit}).

\subsubsection{NOZOMI data}  \label{sec:584n}
In Paper I we used the longitude $258.7^\circ \pm 3.4^\circ$ for the
\HeI\ flow longitude quoted in Section 6 of
\citet{Nakagawa:2008nozomi}.  Referring to the $\pm 3.4^\circ$
uncertainties that we used for the NOZOMI data, LB14 state that the
NOZOMI article contains "... no justification anywhere of such an
uncertainty ...." and this is incorrect.  However, LB14 \emph{are}
correct with their argument that the $\pm 3.4^\circ$ uncertainty is
too small.  We now know that the $\pm 3.4^\circ$ uncertainty given in
the original paper was a typographical error.  The NOZOMI team has
most kindly provided the corrected values for the upwind direction of
the \HeI\ flow: $ \lambda = 258.78^\circ \pm 8.0^\circ, ~ \beta = 3.48
\pm 8.0^\circ$ \citep{Nakagawa:2014erratumnozomi}.  In our refit to
the historical data on the \HeI\ flow direction, we have therefore
adopted these corrected uncertainties for the \HeI\ flow longitude
from the NOZOMI data.  We note that the corrected NOZOMI longitude
affects the statistical likelihood of a constant flow direction over
time using the otherwise same data set as used in Paper I (\S
\ref{sec:refit}).

\section{Conclusions} \label{sec:concl}

While the LB14 speculations about the IBEX and STEREO analyses are
refuted, the differences between the titles of the two papers, "On the
Decades-Long Stability of the Interstellar Wind through the Solar
System" for Lallement \& Bertaux, versus "Decades-Long Changes of the
Interstellar Wind Through Our Solar System" for Frisch et al.,
highlights fundamental questions about the interaction between the
heliosphere and interstellar medium that are not yet understood.

It is shown that small-scale structure, $\le 330$ AU, in the
surrounding LIC may be present.  Structure over these scales is
indicated by the LIC collisional mean free path, possible Alfven waves
propagating along the LIC interstellar magnetic field, and turbulence
related to supersonic motions between the LIC and adjacent clouds that
could drive shocks into the LIC.

In response to the \citet{Frisch:2013sci} meta-analysis of available,
published data at that time, where it was found that the interstellar
wind direction has more likely to have changed by $\sim 6.8^\circ$
over the past forty years than not), \citet{LB14} presented a long
list of arguments that supported their basis for concluding that the
analysis was flawed.  Chief among the incorrect speculations was that
the IBEX-Lo instrumental response function is larger than the measured
value.  The suggestion that azimuthal transport of pickup ions inside
of 1 AU showed that PUIs can only be used as upper limits on the
interstellar wind direction is contradicted by results from the
MESSENGER spacecraft on PUIs inside of of 1 AU.  They also
misunderstood the STEREO analysis procedure.  The arguments of LB14
and our rebuttal on those arguments are summarized in Table
\ref{tab:diff}.

The most valuable comment of LB14 was about the uncertainties on the
NOZOMI \HeI\ wind direction, and it motivated our inquiry to the
NOZOMI team about those uncertainties.  A typographical error in the
original publication \citep{Nakagawa:2008nozomi} has now been
corrected \citep{Nakagawa:2014erratumnozomi} and incorporated into our
analysis (\S \ref{sec:refit}).  A new statistical analysis that
included the corrected NOZOMI uncertainty was performed on the data
used in Paper I.  Again, using the prior sets of observations, but
including this correction, a temporal variation is statistically 
indicated by the data and is
still statistically highly likely.  However, in contrast to the
findings of Paper I, the larger uncertainties of the corrected NOZOMI
longitude now allow a constant flow longitude of $75.0^\circ \pm
0.3^\circ$, although the likelihood of a constant flow is lower than
the likelihood that the flow longitude has varied over time.  The
statistical fit given by header B in Table \ref{tab:refit} should
replace the fit that was provided in Paper I.\footnote{This correction
  brings out the good side of the value of robust scientific
  exchanges.}

The statistical fitting was repeated using improved uncertainties for
the STP 72-1, Mariner 10, and SOLRAD11B results, omitting the Prognoz
6 data and using both of the 2009 and 2010 seasons of IBEX results.
For this fit, the most likely statistical result is found from a
linear fit to the temporal data.  The variation over forty years
corresponds to a longitude variation of $\delta \lambda = 5.6^\circ
\pm 2.4^\circ$.  Again, a constant flow longitude over time was
statistically acceptable but at a lower statistical significance than
the linear variation.  We emphasize that the linear fits are not
equivalent to claiming that the historical variation has actually
occurred as a linear shift with time.

Results from the 2012-2014 IBEX observing seasons of the flow of
interstellar \HeI\ through the heliosphere are now being studied,
however the data contains puzzling effects that suggest that more
complicated processes may contribute to the interaction between
interstellar \HeI\ and the heliosphere
\citep{McComas:2014warm,Leonard:2014isn}.  Those results are not
included in this study.

The comparison of heliospheric data on the interstellar neutral wind
over long time intervals offers the possibility of sampling small
scale structure that likely exists in the surrounding interstellar
material.  There are now over 40 years worth of historical data on the
flow of interstellar \HeI\ through the heliosphere, and not all of it
has been analyzed using contemporary knowledge of the \HeI\ ionization
processes.  Future analyses of these historical data should enrich our
understanding of the interaction between the heliosphere and
interstellar medium.

\begin{acknowledgements}
This work was carried out as a part of the IBEX mission, funded as a
part of NASA’s Explorers Program.  The researchers from SRC PAS were
supported by Polish National Science Center grant 2012-06-M-ST9-00455.
The authors thank Harald Kurcharek for helpful discussions.
\end{acknowledgements}

\appendix
\section{LIC temperature from interstellar absorption lines} \label{app:lic}

Because of the cloudy nature of local ISM, 6\%--33\% of space within
10--15 pc is filled with low-density material that is identified
primarily by cloud velocity \citep{Frisch:2011araa}.
\citet{RLIV:2008} have identified an absorption component
corresponding to the LIC velocity toward $\sim 80$ stars located 2.6
to 74.5 pc away.  Temperatures are reported for many of these stars,
including 15 stars that are within 22 pc (two of which are in a binary
system).  Temperatures are found from the fits to the spectral line
that use an intrinsic Voight profile where the broadening is assumed
to be due both to mass-dependent thermal broadening and a second
mass-independent component.  These temperatures are plotted in
Fig. \ref{fig:licT} against the angle between the star and the LIC
downwind direction.  The reported temperatures range between
$5200^{+1900}_{-1700}$ K and $12050^{+820}_{-790}$ K and show a
tendency for the temperature to decrease toward the downwind
direction.  Alternatively, Redfield and Linsky (2008) also report a
LIC temperature of $7500 \pm 1300$ K using a larger set of 19 stars
out to distances of 68.8 pc.
\citet{Hebrard:1999} reported a LIC temperature of
$8000^{+500}_{-1000}$ K for the binary pair Sirius A and Sirius B,
from the simultaneous fitting of absorption lines from both neutrals
and ions.  The range of the possible LIC temperatures plotted in
Fig. \ref{fig:licT} for the nearest stars indicate either that the gas
temperature varies between sightlines through the LIC, or that other
non-LIC clouds are at LIC-like velocities for some sightlines.

In Paper I we evaluted the LIC temperature from measurements of \FeII,
\MgII, and \CaII\ in the LIC component toward Sirius.  The analysis
was restricted to these refractory species that trace both ionized and
neutral gasses to avoid non-overlapping distributions of neutral and
ionized species because of the ionization gradient in the LIC
\citep{SlavinFrisch:2008}.  A median temperature of
$5800^{+300}_{-700}$ K was found (see the online supplementary
material of Paper I).  The temperature is consistent with the Ulysses
temperature of Witte (2004) and the preferred \HeI\ temperature of
\teminf$= 6300 \pm 390$ K found from IBEX data in
\citet{McComas:2012bow}.  This Sirius temperature was used as an
independent constraint on the temperature uncertainties in the IBEX
parameter tube, which then determined the flow longitude from the IBEX
data that was used in the statistical analysis of Paper I.
Temperatures derived from interstellar absorption lines are not used
in the new fits to the data that are reported in \S \ref{sec:refit}.

\section{The IBEX instrumental corrections and results for the interstellar neutral temperature } \label{app:ibex}

We have determined \teminf\ from the width of the neutral flux
latitude distribution in the analytic method \citet{Moebius:2012isn};
the value obtained through $\chi^2$-minimization by Bzowski et
al. (2012) is also dominated by the latitudinal variation of the
fluxes. The peak height of these distributions, and thus the width,
could possibly be affected by a suppression of the data flow of
individual events through the data system that depends on particle
count rate.  To account for such a potential effect, at that time
still unknown, \citet{Moebius:2012isn} estimated an upper limit in
terms of a maximum event processing time of $\tau < 5$ ms. They
modeled the suspected effect as an instrument dead-time and compared
observations of the same ISN flow at IBEX-Lo energy steps 2 and 4, the
latter with a rate lower by a factor of 5 and thus reduced interface
traffic. Lallement \& Bertaux (2014) have argued that an instrumental
dead-time $\tau \approx 7$ ms would allow a reconciliation of the
angular distributions observed with Ulysses and IBEX, although
\citet{Moebius:2012isn} showed that $\tau \le 5$ ms is an upper limit
and McComas et al. (2012) reported that the effect on the temperature
is much smaller based on follow-up testing and simulation, as
discussed in more detail below.

The telemetry limitation for high-flux neutral atom sources, such as
the ISN flow and the Earth's magnetosphere, was anticipated, and
energy-angle histograms of H and O with $6^\circ$ resolution are
accumulated onboard IBEX to normalize the reported event
rates. Because the width of the IBEX-Lo field-of-view is already
$7^\circ$ FWHM and to eliminate any potentially unanticipated effects
on the data flow by the limited telemetry, the angular resolution used
by Bzowski et al. and \citet{Moebius:2012isn} was restricted to
$6^\circ$

The hardware interface between IBEX-Lo and the data handling system
has a known transfer time of 1.354 ms for each individual event, which
increases for the first $1^\circ$ portion of each $6^\circ$ bin to
2.188 ms because housekeeping data and count rates are transmitted
after each event. Both transmission times are much shorter than the
previously estimated maximum dead-time. In addition, the interface has
a double buffer, which reduces the resulting suppression even
further. The response of the interface to a large variety of event and
rate combinations was tested with a hardware simulator and computed
with Monte Carlo simulations. These tests and reviews of the data
system hardware and software rule out response times longer than the
nominal values quoted above.

It was found that the majority of the event traffic across the
interface, of typical rates of 300-400 s$^{-1}$, originates from an
unanticipated substantial electron flux through the sensor, as
mentioned by \citet{Moebius:2012isn}. Electrons are identified based
on their short time-of-flight and thus are completely eliminated from
ENA signals. However, these events add an almost isotropic (or slowly
varying) base rate that burdens the interface.  Figure \ref{fig:rate}
shows simulation results for the combination of such a base rate and
an angular ISN flow distribution, with a peak rate of 75 events/s (the
maximum ISN event rate observed) as a function of the base event
rate. Shown are the total absolute reduction factor for the base and
peak event rate combination (open symbols), the reduction factor for
the peak relative to the already reduced base rate (solid black), the
resulting increase of the width of the angular distribution (solid
red), and the increase of the deduced temperature (solid blue). While
the total event flow is reduced for the observed base rates of 300-400
events/s by up to 20\%, the relative reduction of the peak rate over
the base is only 2.5-3\%. Only the latter reduction is relevant for
any distortion of the angular distribution, and consequently the width
is increased only by $\sim 1$\%, which translates into an apparent
increase in the deduced temperature of only 2\%. Interestingly, these
effects on the relative height of the peak distribution and the
deduced temperature top out for a base rate of $\sim 700$ events/s at
$<4$\% for the peak rate and $<2.5$\% for the temperature. Any
systematic instrumental effects on the temperature determination from
IBEX observations larger than 2\% and on the peak fluxes as a function
of longitude larger than 2.5\% can be excluded for the observations
analyzed thus far.

It should be noted that the type of events causing this uncertainty
are the unanticipated electrons that are of very little use in the
IBEX ENA analysis, and they have been excluded by a new instrument
mode that started mid-2012.  This operational modification now
eliminates these reduction effects almost entirely. A direct
comparison of angular ISN flow distributions in both modes has
verified in flight the total event rate reduction shown in Figure
\ref{fig:peak} and demonstrated that no significant broadening of the
angular distribution is seen when the base event rate is present.

\section{Separating the temporal and spatial variations in the STEREO data}\label{app:stereo}
The analysis strategy was explained in detail in \citet{Drews:2012}.
D12 disentangle the temporal and spatial variations in the STEREO data
through the use of equation 3 in their paper, where the pickup ion
count rates are separated into the terms representing the
time-independent production model for the pickup ions, $V_\mathrm{r}
(\phi)$ for the azimuthal angle $\phi$, and the time-dependent
modulation parameter $M\mathrm(t)$, which describes the fluctuations
of pickup ion count rates induced by solar wind variabilities on short
time scales.  Under the assumption that the production mechanism is
the same for each orbit, an approximation of the modulation parameter,
$G\mathrm(t)$, can be determined from the variability of the pickup
ion count rates with respect to the average of all of the orbits, as
is described through equation 4 in their paper. The possible effects
of short-term variabilities of high-speed solar wind streams or
electron impact ionization were overcome by fitting the product of the
time-dependent modulation parameter, $G\mathrm(t)$, and a model for
the angular distribution of interstellar pickup ions $V_\mathrm{m}
(\phi,\lambda)$ to the angular distribution of He$^+$, O$^+$, and
Ne$^+$ observed by STEREO/PLASTIC (equation 6 in their paper). Not
including $G\mathrm(t)$ into the fitting procedure, i.e.,
$G\mathrm(t)=\rm{constant}$, would mean disregarding the temporal
fluctuations of pickup ion count rates induced by solar wind
variabilities such as electron impact ionization, and is therefore
comparable to a standard fitting technique, i.e., fitting a Gaussian
distribution to the focusing cone and the pickup ion crescent.

LB14 have also objected to the method of comparing data between two
separate orbits using different possible permutations of the four
orbits, arguing that the different orbit combinations ``are not fully
independent''.  We disagree with this viewpoint.  Due to the random
nature of the pickup ions fluctuations in each orbit, the intrinsic
comparisons between pairs of different orbits produces independent
information.  The dependent variable that arises from these different
permutations of orbit comparisons is the invariant component of the
pickup ion distribution that does not change with time and depends
only on the inflow direction of the interstellar helium atoms. The
independent variable is the estimate of the relative modulation
parameter $G\mathrm(t)$, which was deduced from equation 4 in their
paper and changes for each permutation of orbit combinations. Because
$G\mathrm(t)$ is a key parameter to recover the inflow direction from
pickup ion data, the intercomparison of the individual orbits is
required to determine the uncertainty of the modulation parameter,
$\Delta G\mathrm(t)$, which in turn is necessary to determine a
scientifically justified estimation of the uncertainty of the inflow
direction of interstellar matter, $\Delta \lambda$.

Further simulations of the STEREO data are presented in
\citet{Drews:2013phd}.  This extension of the statistical analysis
compares two different approaches, the numerical co-adding of the
count rates that ignores the time variability of the STEREO data,
versus simulations of the properties of each orbit compared to the
overall data sample.  These comparisons were performed for several
different sets of conditions and are presented in Drews (2013),
chapter 4.3. The statistical approach was based on 1500 randomly
generated data sets using four different kind of modulations.  Drews
compared the results of the simulations to those of a standard fitting
technique.  \citet[][Fig. 4.4]{Drews:2013phd} showed that the standard
fitting technique provides a less precise fit to the STEREO data,
producing an uncertainty for the focusing cone and crescent position
that is $\sim$2 times larger than the uncertainty found using the full
statistical analysis.

\begin{deluxetable}{llll} 
\tablecolumns{8} 
\tablewidth{0pc} 
\tabletypesize{\footnotesize}
\tablecaption{Comments on \citet{LB14} criticisms of data and usage in Paper I \tablenotemark{A} \label{tab:diff}}
\tablehead{ 
\colhead{}&\colhead{Argument} &  \colhead{Comment} & \colhead{Section} \\
\colhead{}&\colhead{} &  \colhead{} & \colhead{(Appendix)}}
\startdata 
&\multicolumn{2}{l}{\it \underline{Criticisms regarding IBEX analysis:}} &   \\ 
1& Ulysses and IBEX velocity-longitude & True for hot models using Maxwellian &   \ref{sec:ibexfct} \ref{sec:ibexseason}  (\ref{app:ibex}) \\
 & combinations give same perihelion & distributions, but irrelevant to results &  \\
2& IBEX-Lo instrumental response & Incorrect; the IRC, dead-time correction, & \ref{sec:ibexirf} (\ref{app:ibex}) \\
 &  correction (IRC) is wrong &   and data throughput are well understood   &  \\ 
3& Correct IRC makes IBEX & Incorrect; the IBEX results rely on & \ref{sec:ibexfct} \\
 & and Ulysses longitudes agree & variations of the peak as a function of &  \\
 & & latitude and observer longitude   &  \\
4& Warsaw modeling based on & Incorrect; Warsaw models compared to  & \ref{sec:ibexirf} \\
 &wrong IRC corrections  & data based on a correct and tested IRC  &  \\
&\multicolumn{2}{l}{\it \underline{Criticisms regarding STEREO analysis:}}  &  \\
5&Short-term electron-impact ionization  & Incorrect; short-term EII is distinguished  & \ref{sec:stereo}, \ref{sec:stereotime} \\ 
 & (EII) not included in  data analysis & from long-term ionizations in methodology  &  \\
6& Orbit groupings are not & Incorrect; each orbit pairing selects  & \ref{sec:stereo} (\ref{app:stereo}) \\
& statistically independent & invariant distributions from comparisons &  \\
& & of data acquired over separate time-spans  &  \\
7&Simulations of uncertainties of & Uncertainty simulations are   & \ref{sec:stereo} (\ref{app:stereo}) \\
 &  STEREO data are incorrect&   appropriate for variable PUI data  &  \\
8&Secondary oxygen invalidates usage & Incorrect; secondary oxygen population  & \ref{sec:stereosec} \\
 &  of oxygen PUIs & minor compared to primary population, and &  \\
 &  & if included push longitudes into the &  \\
 &  & opposite direction than suggested &  \\
9& Longitudes from PUI focusing cone & Maybe, but transport effects appear to be  & \ref{sec:puitransport} \\ 
 &  and upwind crescent are upper limits & insignificant, based on comparisons between  &  \\ 
 &  due to transport effects inside 1 AU & several different data sets &  \\
&\multicolumn{2}{l}{\it \underline{Criticisms regarding use of Ultraviolet Data:}}  &  \\
10& Mariner 10 longitude is based on & Incorrect: Mariner 10 papers quote  & \ref{sec:584m} \\
 &  both \HI\ and \HeI\ & longitude of \HeI\ wind separately  &  \\
11&\HeI\ flow direction from Prognoz 6 &  The Prognoz 6 \HeI\ direction from geometric  & \ref{sec:584p} \\
&  geometric analysis not valid & arguments has uncertainties and was included  &  \\
&  & since it was not repudiated in the literature  &  \\
12& NOZOMI uncertainties of $\pm 3.4^\circ$ are & True; the results of the 2014 NOZOMI  & \ref{sec:584n} \\
 &  too small & erratum are used in this paper  &  \\
\enddata 
\tablenotetext{A}{Numbers in last column show section or appendix where the topic is discussed.}
\end{deluxetable} 

\begin{deluxetable}{clccc}
\rotate
\tabletypesize{\footnotesize}
\tablecaption{Fits to historical interstellar wind data (\today) \label{tab:refit}}
\tablewidth{0pt}
\tablehead{
\colhead{Num.} & \colhead{Basis}   & \colhead{Fit\tablenotemark{*}} & \colhead{Reduced} & \colhead{p-value}  \\
\colhead{} & \colhead{of fit}   & \colhead{$\lambda$(deg.)} & \colhead{$\chi^2$} & \colhead{} 
}
\startdata
                 
\multicolumn{5}{l}{\hspace*{0.08in} \it A.  Paper I:\tablenotemark{A}:} \\
1   &  Linear &  $ 70.6 (\pm 1.6) + 0.17 (\pm 0.06)* t_\mathrm{1970} $  &  0.97 & 0.49 \\
2   &  Constant $\lambda$ &  $ 75.1 (\pm 1.3) $  &  1.71 & 0.031 \\

\multicolumn{5}{l}{\hspace*{0.08in} \it B.  Corrected NOZOMI uncertainties:\tablenotemark{B}} \\
3   &  Linear & $ 71.3 (\pm 1.6) + 0.14 (\pm 0.06)* t_\mathrm{1970} $ &  0.7\tablenotemark{C} & 0.82 \\
4   &  Constant $\lambda$ & $ 75.0 (\pm 0.3) $ &  1.1 & 0.35 \\

\multicolumn{5}{l}{\hspace*{0.08in} \it C.  Two IBEX data points, modified 584\AA\ uncertainties:\tablenotemark{D}} \\
5   &  Linear & $ 71.4 (\pm 1.6) + 0.14 (\pm 0.06)* t_\mathrm{1970} $  & 0.8 & 0.3 \\
6   &  Constant $\lambda$ & $ 75.0 \pm 0.3$ & 1.3 & 0.2 \\
7   &  Parabolic $\lambda$ fit & $ 74.1 (\pm 1.4)  - 0.11 (\pm 0.11)* t_\mathrm{1970}  + 5.0 (\pm 2.1) 10^{-3}*t_\mathrm{1970}^2 $ & 0.6 & 0.14 \\
 \enddata

\tablenotetext{*}{$\lambda$ is the ecliptic longitude of the interstellar
wind and $t_\mathrm{1970}$ is the elapsed time since 1970.  A p-value larger than 0.05 passes the p-test
and therefore is not unlikely.}
\tablenotetext{A}{This fit is the original linear fit in Paper I that used the NOZOMI value from \citet{Nakagawa:2008nozomi}.}
\tablenotetext{B}{These fits used the historical wind data in Paper I, except that the
uncertainties for the NOZOMI value are corrected to incorporate the erratum of Nakagawa et al. (2014).  
This linear fit should replace the fit listed in Paper I.  See Section \ref{sec:584}.}
\tablenotetext{C}{This fit is accepted and is more likely than fit number 4.  The small value
of $\chi^2$ suggests that the uncertainties may be overestimated by $\sim 20$\%.}
\tablenotetext{D}{These fits incorporate uncertainties for the wind longitude for the STP2-1, Mariner 10, and SOLRAD11B data
that include only the longitude uncertainties (see Section \ref{sec:584}). The geometric Prognoz 6 direction is
omitted from the fit according to the discussion in Section \ref{sec:584}.
The \HeI\ wind directions from the individual IBEX 2009 and 2010 observing seasons are included independently 
\citep[Sections \ref{sec:ibexseason}, \ref{sec:refit}, and see][]{Bzowski:2012isn}.  }
\end{deluxetable}
\pagebreak
\newpage

\newpage

\newpage

\newpage

\newpage

\newpage
\begin{figure}
\plotone{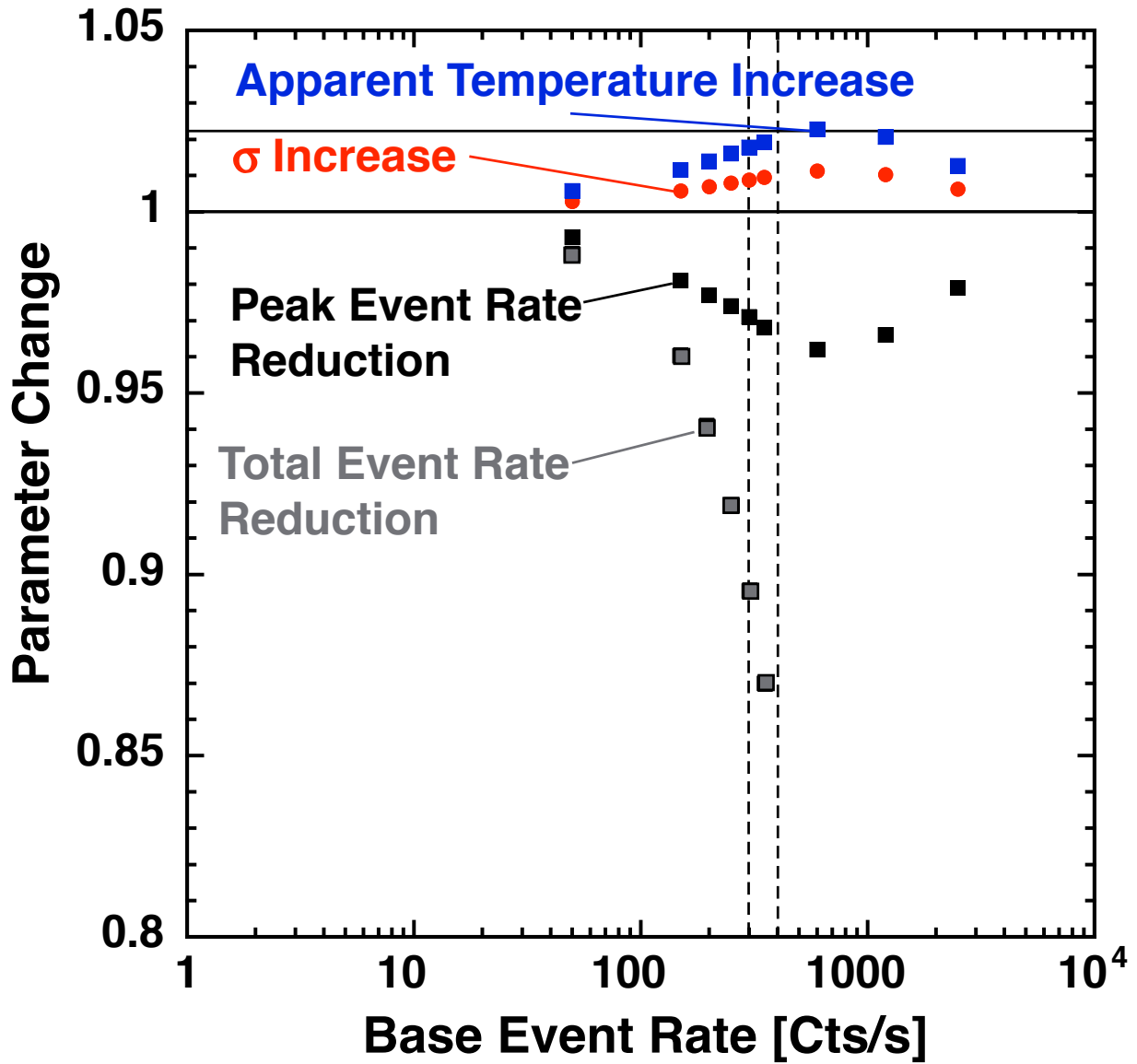}
\caption{Simulated effects on the observed flux distribution of the
  ISN flow for the maximum peak rate of 75 events/s. Absolute total
  rate reduction (open symbols), reduction of the ISN flow peak rate
  relative to the base rate reduction (solid black), increase of the
  angular width $\sigma$ in latitude (solid red), and resulting
  apparent temperature increase (solid blue) as a function of
  different continuous base event rates.  }\label{fig:rate}
\end{figure}

\newpage
\begin{figure}
\begin{center}
\plotone{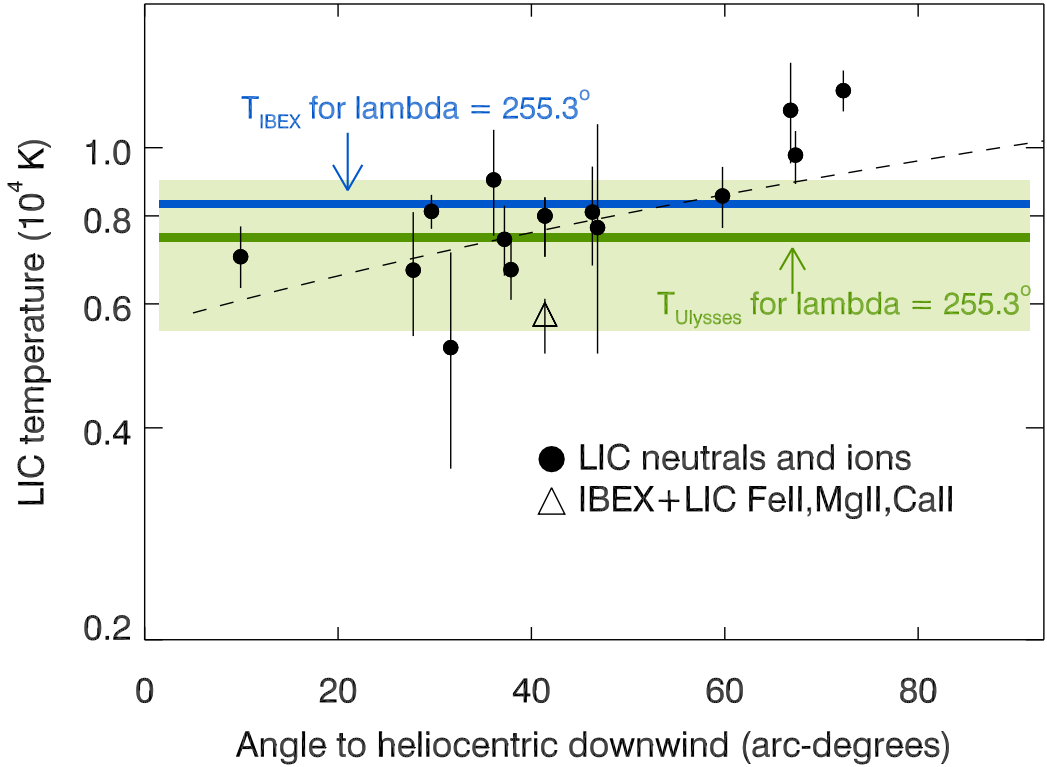}
\end{center}
\caption{The temperatures of the LIC toward stars within 22 pc that
  sample the LIC, according to Table 1 of \citet[][dots]{RLIV:2008},
  are plotted against the angle between the star and the ISN downwind
  direction ($\ell = 184^\circ,~b=-12^\circ$ in galactic coordinates).
The triangle shows the LIC temperature from Paper I, which was
    determined from the IBEX parameter range combined with \FeII,
    \MgII, and \CaII\ absorption line data in the LIC component
  toward Sirius.  The wide green shaded region shows the uncertainty
  on the temperature range found from the reanalysis of the Ulysses
  data by \citet{Bzowski:2014ulysses}.  The blue and dark-green bars
  show, respectively, the temperatures required by IBEX and Ulysses
  data if the true interstellar flow upwind direction is the best-fit
  value found from the Ulysses data, or 255.3\deeg\ according to
  \citet{Bzowski:2014ulysses}.  The difference between these two
  temperatures (853 K) shows that IBEX and Ulysses data are not in
  agreement.  The dashed line shows a second order fit through the
  interstellar temperature data (dots), and is meant to guide the eye
  and illustrate the systematic decrease in reported cloud temperature
  as the angle to the downwind direction decreases.  If this apparent
  temperature decrease is spurious, it could represent the
  misidentification of LIC components in sidewind directions.
}\label{fig:licT}
\end{figure}

\end{document}